\documentclass[
superscriptaddress,
preprintnumbers,
twocolumn,
nofootinbib,
amsmath,amssymb,
 aps,eqsecnum,
]{revtex4-2}

\usepackage{graphicx}
\usepackage{dcolumn}
\usepackage{bm}
\usepackage{hyperref}

\usepackage{comment}
\usepackage{caption}
\usepackage{subcaption}
\usepackage{booktabs}
\usepackage{color}
\usepackage{mathtools}
\usepackage[normalem]{ulem} 
\usepackage{amsmath}
\usepackage{orcidlink}

\usepackage{url}
\hypersetup{colorlinks=true, citecolor=cyan}

\newcommand{\be}{\begin{equation}}
\newcommand{\ee}{\end{equation}}

\definecolor{mygreen}{RGB}{0,130,0}

\begin{document}
\preprint{RESCEU-32/25}

\author{Daiki Watarai
\orcidlink{0009-0002-7569-5823}}
\affiliation{Graduate School of Science, The University of Tokyo, Tokyo 113-0033, Japan}
\affiliation{Research Center for the Early Universe (RESCEU), Graduate School of Science, The University of Tokyo, Tokyo 113-0033, Japan}
\author{Kent Yagi
\orcidlink{0000-0002-0642-5363}} 
\affiliation{Department of Physics, University of Virginia, P.O. Box 400714, 382 McCormick Road, Charlottesville, Virginia 22904-4714, USA}
\author{Shammi Tahura
\orcidlink{0000-0001-5678-5028}}
\affiliation{Department of Physics and Astronomy, University of Iowa, Iowa City, IA 52242, USA}

\newcommand*{\diff}{\,\mathrm{d}}
\newcommand*{\isco}{\mathrm{ISCO}}
\newcommand*{\adins}{adiabatic inspiral}
\newcommand*{\ins}{inspiral}
\newcommand*{\tra}{transition}
\newcommand*{\plu}{plunge}
\newcommand*{\Ecirc}{\hat{E}^\mathrm{(circ)}}
\newcommand*{\Lcirc}{\hat{L}^\mathrm{(circ)}}
\newcommand*{\Omegacirc}{\Omega^\mathrm{(circ)}}

\date{\today}
\begin{abstract}

Binary black hole mergers with asymmetric component masses are key targets for both third-generation ground-based and future space-based gravitational-wave (GW) detectors, offering unique access to the strong-field dynamics of gravity. The evolution is commonly divided into three stages: the adiabatic inspiral, the transition, and the plunge. To date, constructions of inspiral-transition-plunge waveforms have largely focused on Schwarzschild or Kerr background spacetimes. In this paper, we extend these efforts to spacetimes beyond Kerr by constructing such waveforms in a Kerr--Newman background. For simplicity, we allow the primary black hole to carry spin and charge while keeping the secondary object neutral and non-spinning. We work in the small charge-to-mass ratio regime and adopt the Dudley--Finley approximation, in which the gravitational and electromagnetic perturbations decouple. In particular, the gravitational sector satisfies a Teukolsky-like equation, enabling only minimal modifications relative to the Kerr case when constructing the waveform. Having the inspiral-transtion-plunge waveforms in hand, we studied observational prospects for constraining the charge of the central black hole. We find that, for intermediate–mass–ratio mergers observed with the Einstein Telescope, explicitly modeling the post-inspiral dynamics significantly tightens charge-to-mass ratio constraints. In particular, the bounds on the charge-to-mass ratio can reach $\mathcal{O}(10^{-3})$ in the region of primary masses and spins where the post-inspiral signal dominates, yielding charge bounds that can be orders of magnitude tighter than those obtained from the inspiral alone or from the current bound with GW150914. These results lay the groundwork for inspiral-transition-plunge waveform modeling in beyond-Kerr spacetimes and for probing non-Kerr signatures in future GW observations.

\end{abstract}

\title{\textbf{Inspiral–Transition–Plunge Gravitational Waveforms Beyond Kerr:\\ A Kerr–Newman Case Study}
}
\maketitle

\section{Introduction}
\label{sec:intro}
Gravitational-wave (GW) astronomy now offers an unprecedented avenue for directly probing the strong-field regime of gravity.
Over the past decade, the LIGO–Virgo–KAGRA (LVK) Collaboration has detected more than two hundred GW events from binary black hole coalescences~\cite{LIGOScientific:2016aoc, LIGOScientific:2016vlm, LIGOScientific:2018mvr, LIGOScientific:2020ibl, KAGRA:2021vkt, LIGOScientific:2025slb}. 
These data provide a unique opportunity to test general relativity (GR), the standard theory of gravity, in the strong-gravity and near-horizon regimes.
While a wide range of tests of GR have been performed, none have revealed statistically significant deviations within current observational precision~\cite{LIGOScientific:2016vlm,Yunes:2016jcc, LIGOScientific:2019fpa, LIGOScientific:2020tif, LIGOScientific:2021sio, KAGRA:2025oiz, LIGOScientific:2025obp,Yunes:2025xwp,Gupta:2025utd}.
In the coming decades, third-generation ground-based GW observatories such as the Einstein Telescope (ET)~\cite{Hild:2010id, Punturo:2010zz} and Cosmic Explorer (CE)~\cite{Reitze:2019iox}, as well as the space-based GW interferometer LISA~\cite{Baker:2019nia}, are expetected to operate and detect signals with much higher signal-to-noise ratios (SNRs), enabling far more precise tests of possible deviations from GR.

To extract properties of the source, it is necessary to construct an accurate model for the incoming GW signal to compare against the data.
For GW signals detected so far, i.e., comparable-mass binary coalescences, various types of waveform models have been proposed and applied to data analysis.
Most of them, such as effective-one-body models (e.g.~\cite{Buonanno:1998gg, Bohe:2016gbl, Cotesta:2018fcv}) and inspiral-merger-ringdown phenomenological models (e.g.~\cite{Ajith:2007qp, Khan:2015jqa, Pratten:2020ceb, Colleoni:2024knd}), are constructed by using analytic results based on the perturbation theory frameworks, the post-Newtonian approximation for the inspiral and the BH perturbation theory for the ringdown, and numerical relativity (NR) simulation data for the merger phase.

In addition to comparable-mass binary coalescences, future detectors are expected to observe asymmetric-mass systems, in particular binaries with extreme and intermediate mass ratios, typically, the smaller-to-larger mass ratios of $10^{-6}$–$10^{-4}$ and $10^{-4}$–$10^{-2}$, respectively.
For such systems, a suitable theoretical framework to consistently treat the whole coalescence process is the BH perturbation theory, where the larger BH acts as the background spacetime while the smaller body moves on and perturbs this geometry.
Indeed, the post-Newtonian (PN) approximation ceases to be accurate near the final stages of the binary evolution, where the smaller object's speed approaches a fraction of the speed of light and the object becomes highly relativistic.
On the other hand, NR simulations become prohibitively expensive, as resolving the small body requires extremely fine grid spacing and leads to rapidly increasing computational cost.

Asymmetric-mass binaries are promising systems for providing an excellent opportunity to test the nature of BHs. 
For example, extreme–mass–ratio inspirals (EMRIs) provide exceptionally clean tests owing to their long inspiral duration and well-controlled perturbative description.
In addition, intermediate–mass–ratio mergers will be intriguing targets for upcoming ground- and space-based detectors, as both their inspiral and post-inspiral signals can be observed with high SNRs.
The inspiral stage of such systems, so-called intermediate mass ratio inspirals, encompasses a large number of orbital cycles within the detector band, allowing for high-precision parameter estimation.
In addition, one of the authors showed that higher-harmonic quasi-normal modes are efficiently excited in cases involving astrophysically relevant primary BHs with high spins~\cite{Watarai:2024huy}, enabling precise estimation of source properties even from a ringdown analysis alone~\cite{Watarai:2024vni}.
These characteristics make intermediate–mass–ratio mergers ideal targets for robust searches for beyond-GR or beyond-Kerr signatures by comparing the parameters independently inferred from the inspiral, post-inspiral, and full data sets.

The coalescence of an asymmetric-mass binary BH system is commonly divided into three stages: the adiabatic inspiral, transition, and plunge.
In the adiabatic inspiral, the orbital radius evolves due to GW emission on a timescale much longer than the orbital period.
Thus, the motion can be treated as a sequence of quasi-circular orbits, each specified by the conserved energy and angular momentum at that radius.
As the smaller BH approaches the innermost stable circular orbit (ISCO) of the larger BH, the adiabatic approximation gradually breaks down.
The radial motion begins to accelerate under the combined effects of radiation reaction and strong-field gravity, defining the transition regime.
In the final plunge phase, the radiation-reaction timescale becomes much shorter than the orbital period, and the motion becomes well approximated by a geodesic as the smaller BH rapidly plunges into the larger one.

Apte and Hughes~\cite{Apte:2019txp}, hereafter AH19, proposed a prescription for consistently stitching together these three regimes and computing inspiral–transition–plunge gravitational waveforms in the Kerr background.
Their key idea is to construct a single continuous worldline for the smaller body's trajectory by identifying the time intervals in which the solutions to the equations of motion in each regime closely agree, and then smoothly matching them at appropriate collocation times.
One can then substitute this wordline into the source term of the time-domain Teukolsky equation, enabling the computation of the corresponding numerical gravitational waveforms.
Such a complete waveform is also important for data analysis, because restricting the analysis to only a portion of the signal, whether one focuses solely on the inspiral, the plunge, or any other truncated segment, can discard physical information contained in the full waveform and lead to systematic biases in the inferred parameters~\cite{Mandel:2014tca}.

So far, the construction of inspiral-transition-plunge waveforms for asymmetric-mass binaries has focused on either the Schwarzschild or Kerr BH background. 
It is important to go beyond the current setup when testing beyond-GR or beyond-Kerr effects. 
To take a first step towards this goal, the purpose of this paper is to compute, for the first time, the inspiral-transition-plunge waveforms in beyond-Kerr spacetimes by extending Ref.~\cite{Apte:2019txp}.
In this work, we take the Kerr--Newman spacetime as an example.
The Kerr--Newman spacetime is known as the unique stationary BH solution to the four-dimensional Einstein–Maxwell equation, characterized by its mass, spin angular momentum, and electric charge.
Although astrophysical BHs are generally expected to be nearly neutral and hence well approximated by the Kerr solution, several mechanisms motivate one to consider charged BHs as physically relevant extensions of the Kerr paradigm.
Even within the standard model, processes such as the Wald mechanism can induce a small charge on a spinning BH in an external magnetic field~\cite{Wald:1974np}, while theories beyond the standard model often predict BHs carrying additional gauge charges—for example through hidden U(1) sectors (e.g.~\cite{Ackerman:2008kmp, Feng:2009mn, Foot:2014osa, Foot:2014uba, Agrawal:2016quu, Cardoso:2016olt}) or vector–tensor extensions of gravity~\cite{Moffat:2005si, Moffat:2016gkd}.
These scenarios, many of which are considered candidates for resolving the dark matter problem, establish the Kerr--Newman spacetime as a natural and theoretically well-motivated setting for exploring possible deviations from the Kerr metric.

\begin{figure*}
    \centering
    \includegraphics[width=\linewidth]{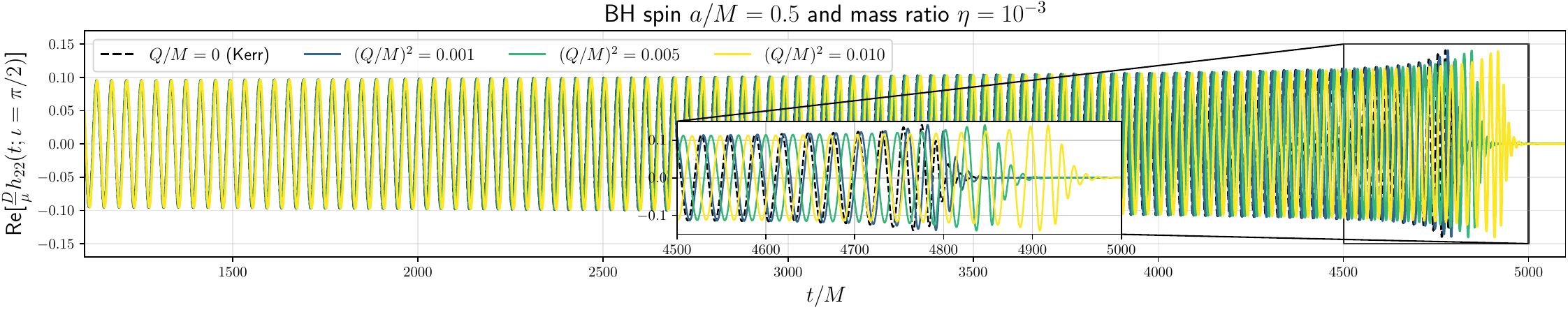}
    \caption{
    Gravitational waveforms of the $(\ell,m)=(2,2)$ mode are shown for intermediate–mass–ratio binary mergers with the mass ratio $\eta = 10^{-3}$, evaluated at an inclination angle $\iota = \pi/2$, for the BH spin $a/M = 0.5$.
    The black dashed curve represents the Kerr case ($Q/M = 0$), while the colored solid curves correspond to $(Q/M)^2 = 0.001$ (navy), $0.005$ (green), and $0.01$ (yellow). 
    }
    \label{fig:h22_intro}
\end{figure*}

\begin{figure}
    \centering
    \includegraphics[width=\linewidth]{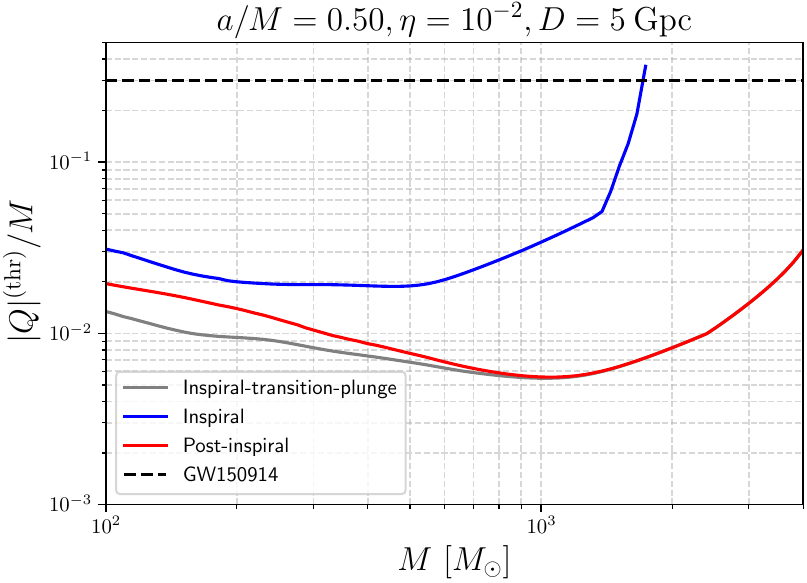}
    \caption{Distinguishable charge-to-mass ratio $|Q|^\mathrm{(thr)}/M$ using the inspiral-transition-plunge, inspiral, and post-inspiral waveforms with ET.
    The horizontal black dashed line indicates the threshold charge for GW150914 $|Q|/M = 0.3$~\cite{Bozzola:2020mjx, Carullo:2021oxn}, shown for reference.
    }
    \label{fig:Q_thr_intro}
\end{figure}

\subsection{Overview of the computational framework}
\label{sec:overview_computation}
In this subsection, we summarize the theoretical and computational framework employed to construct inspiral–transition–plunge waveforms in the Kerr--Newman spacetime.

\paragraph*{\textbf{Secondary object.}}
Throughout this work, we model the smaller body as a non-charged, non-spinning test particle.  
For a neutral secondary, there is no direct electromagnetic radiation sourced by the particle itself.  
In the Kerr--Newman background, gravitational and electromagnetic perturbations are coupled, which, in principle, generates an indirect electromagnetic response. 
However, this induced component is expected to be subdominant compared to the leading gravitational emission and is thus ignored in the present analysis. 

\paragraph*{\textbf{Teukolsky formalism and the Dudley--Finley approximation.}}
Because we freeze the electromagnetic perturbation to zero, the equation for the gravitational perturbation reduces to the Dudley--Finley equation, which provides a separable approximation to the full Kerr--Newman perturbation system.  
To compute the emitted gravitational radiation, we therefore adopt the frequency-domain Teukolsky framework together with the Dudley--Finley approximation.
Although the Dudley--Finley equation is not the exact gravitational
perturbation equation for Kerr--Newman BHs, recent studies have shown
that the quasi-normal mode spectrum within the approximation closely matches that of the full calculation, particularly for small and moderate values of the charge-to-mass ratio~\cite{Saha:2025nsg}.  
These results support the use of the Dudley--Finley equation as a physically reliable and computationally efficient tool for modeling GW emission from perturbed Kerr--Newman BHs.

\paragraph*{\textbf{Construction of the particle worldline.}}
The orbital evolution consists of three stages: adiabatic inspiral, transition, and plunge.  
We follow the prescription of AH19 to construct the transition solution and match it smoothly from the adiabatic inspiral to the final geodesic plunge.
This procedure yields a continuous worldline suitable to generate the waveforms for the whole coalescence process.

\subsection{Summary of our results}
\label{sec:intro_summary}
Before presenting the detailed formalism and results, we summarize the main findings of this work.  
Figure~\ref{fig:h22_intro} illustrates the impact of the BH charge on the gravitational waveform for an intermediate-mass-ratio binary with the mass ratio $\eta = 10^{-3}$ and the primary mass $M$ and spin $a/M = 0.5$ throughout the inspiral, transition and plunge.
Even for small values of the charge, $(Q/M)^2 \sim 10^{-3}$--$10^{-2}$, the waveform exhibits a clear deviation from the Kerr prediction in the phase, especially during the late inspiral and ringdown.  

To assess the measurability of the BH charge with ET, we compute the threshold charge-to-mass ratio $|Q|^{\mathrm{(thr)}}/M$ based on the waveform mismatch, comparing results with the inspiral wavform alone, the post-inspiral waveform only, and the full inspiral-transition-plunge waveform.
Figure~\ref{fig:Q_thr_intro} shows the resulting thresholds for $a/M = 0.5$, $\eta = 10^{-2}$, and the luminosity distance of $D = 5~\mathrm{Gpc}$.  
For systems with $M \sim 10^{2}$--$10^{3} M_{\odot}$, the charge can be constrained down to the level of $|Q|/M \sim {\rm few} \times 10^{-3}$ if we have a model for the post-inspiral part, while $|Q|/M \sim {\rm few} \times 10^{-2}$ from the inspiral alone.
In addition, the complete inspiral-transition-plunge waveform provides the most robust constraints across the entire mass range.  
The result demonstrates that the inclusion of post-inspiral dynamics can lead to substantially more precise measurements of the BH charge, which would be inaccessible using inspiral-only waveform models.

\subsection{Organization of this paper}
This paper is structured as follows. 
Section~\ref{sec:KN_BH} provides a brief introduction to the Kerr--Newman geometry and the motion of a test particle. 
In Sec.~\ref{sec:Teukolsky_formalism}, we summarize the Teukolsky formalism used for GW calculations. 
Section~\ref{sec:ITP_GWs} presents the construction of the complete worldline for an asymmetric-mass merger and the corresponding waveforms. 
In Sec.~\ref{sec:observation_prospect}, we explore the observational prospects enabled by these waveforms: we evaluate how modeling the post-inspiral dynamics improves constraints on the BH charge in intermediate–mass–ratio mergers detectable by ET, and we assess the sensitivity of extreme–mass–ratio inspirals observed by space-based interferometers to tiny deviations from the Kerr geometry. 
Finally, in Sec.~\ref{sec:discussion} we discuss the assumptions underlying our calculations and outline several directions for future work, and in Sec.~\ref{sec:conclusion} we summarize our findings.
We use the geometric units of $c=G=1$ throughout sections up to Sec.~\ref{sec:observation_prospect}, and restore the explicit dependence on these constants thereafter.

\section{Test particle motion around the Kerr--Newman black hole}
\label{sec:KN_BH}

In this section, we first introduce the spacetime metric for a Kerr-Newman BH. We then present the geodesic equations and show key quantities for a test particle in a circular motion around a Kerr-Newman BH.

\subsection{Spacetime metric}
We begin by presenting the spacetime metric for a Kerr--Newman BH, which is given by~\cite{Newman:1965my}
\begin{equation}
\label{eq:metric}
\begin{split}
    \diff s^2 = &-\frac{\Delta}{\Sigma} \left( \diff t -a\sin^2{\theta} \diff \phi \right)^2 + \Sigma \left( \frac{\diff r^2}{\Delta} + \diff \theta^2 \right)\\ 
    &+ \frac{\sin^2 \theta}{\Sigma}\left[ a\diff t - (r^2+a^2)\diff \phi \right]^2\:,
\end{split}
\end{equation}
with 
\begin{align}
    \Delta &:= r^2 - 2Mr + a^2 + Q^2\:, \label{eq:Delta} \\
    \Sigma &:= r^2 + a^2 \cos^2\theta\:,
\end{align}
where $M$ is the BH mass, $a$ is the spin parameter, and $Q$ is the charge. 

The event horizons, defined by $\Delta=0$, are located at
\begin{equation}
    r_\pm = M \pm \sqrt{M^2 - (a^2 + Q^2)}\:,
\end{equation}
where $\pm$ denote the outer and inner horizons respectively.
The condition for the existence of the horizons is
\begin{equation}
    M^2 > a^2 + Q^2\:.
\end{equation}

This spacetime is stationary and axisymmetric, meaning that two Killing vectors exist, $(\partial/\partial t)^\mu$ and $(\partial/\partial \phi)^\mu$. 
Accordingly, the test particle's energy and angular momentum, $E$ and $L$, are conserved along a geodesic.
We also define the normalized quantities $\hat{E}=E/\mu$ and $\hat{L}=L/(\mu M)$, where $\mu$ is the particle's rest mass.

\subsection{Circular orbit}
The geodesic equations for the Kerr--Newman spacetime can be expressed as
\begin{align}
    \frac{\diff t}{\diff \lambda} &= -a(a\hat{E}\sin^2\theta-\hat{L}) + \frac{r^2+a^2}{\Delta}P\:,\label{eq:t_geo}\\
    \frac{\diff r}{\diff \lambda} &= \pm\sqrt{R}\:,\label{eq:r_geo}\\
    \frac{\diff \theta}{\diff \lambda} &= \pm \sqrt{\Theta}\:,\label{eq]theta_geo}\\
    \frac{\diff \phi}{\diff \lambda} &= -\left( a\hat{E}-\frac{\hat{L}}{\sin^2\theta} \right) + \frac{a}{\Delta}P\:,\label{eq:phi_geo}
\end{align}
where
\begin{align}
    P(r, \hat{E}, \hat{L}) &= \hat{E} (r^2+a^2) - a\hat{L}\:,\\
    R(r, \hat{E}, \hat{L}) &= P^2 - \Delta\left[r^2 + (\hat{L}-a\hat{E})^2 + C \right]\:,\\
    \Theta(r, \hat{E}, \hat{L}) &= C - \cos^2\theta \left[ a^2(1-\hat{E}^2) + \frac{\hat{L}^2}{\sin^2\theta} \right]\:,
\end{align}
with $C$ representing the Carter constant while $\lambda$ being the Mino time defined by $\diff \lambda = \diff \tau/\Sigma$ with $\tau$ representing the proper time of the test particle in geodesic motion.
In this paper, we restrict our analysis to equatorial motion ($\theta=\pi/2$) and set $C=0$.

Circular orbits are determined by the conditions $R = 0$ and $\partial R /\partial r =0$. Solving these, the energy and angular momentum of the particle in a circular orbit with radius $r$ are given by
\begin{align}
    \hat{E}^{\mathrm{(circ)}}(r) &= 
    \frac{a\sqrt{M r - Q^2} + (Q^2 + r^2 - 2Mr)}
    {r\sqrt{2Q^2 + r^2 - 3Mr + 2a\sqrt{M r - Q^2}}}\:, \label{eq:E_circ_KN}\\[1ex]
    \hat{L}^{\mathrm{(circ)}}(r) &= 
    \frac{a(Q^2 - 2Mr) + (a^2 + r^2)\sqrt{M r - Q^2}}
    {r\sqrt{2Q^2 + r^2 - 3Mr + 2a\sqrt{M r - Q^2}}}\:.\label{eq:L_circ_KN}
\end{align}
The orbital angular velocity then reads
\begin{equation}
\label{eq:omega_circ_KN}
    \Omega^{\mathrm{(circ)}}(r) := \frac{\diff \phi}{\diff t}(r) 
    = \pm\frac{\sqrt{M r - Q^2}}{r^2 \pm a\sqrt{M r - Q^2}}\:. 
\end{equation}
The upper (lower) signs correspond to prograde (retrograde) motion, respectively. 

The innermost stable circular orbit (ISCO) radius is determined by further imposing $\partial^2 R/\partial r^2 = 0$, leading to the condition~\cite{Liu:2017fjx}
\begin{widetext}
\begin{equation}
    M r_\mathrm{ISCO}\left(6M r_\mathrm{ISCO} - r_\mathrm{ISCO}^2 - 9Q^2 + 3a^2\right) 
    + 4Q^2(Q^2 - a^2) - 8a\,(M r_\mathrm{ISCO} - Q^2)^{3/2} = 0\:. 
    \label{eq:ISCO_condition_KN}
\end{equation}
\end{widetext}
$r_\mathrm{ISCO}$ is obtained by solving Eq.~\eqref{eq:ISCO_condition_KN} numerically, and the corresponding $\hat{E}_\mathrm{ISCO}$, $\hat{L}_\mathrm{ISCO}$, and $\Omega_\mathrm{ISCO}$ follow by substituting $r=r_\mathrm{ISCO}$ into Eqs.~\eqref{eq:E_circ_KN}–\eqref{eq:omega_circ_KN}.

\section{The Teukolsky formalism}
\label{sec:Teukolsky_formalism}
In this work, we consider perturbations of a slightly charged, rotating BH, i.e., a Kerr--Newman background with $|Q|/M \ll 1$, sourced by a neutral, non-spinning point particle.
Under such a setup, no electromagnetic radiation is emitted from the motion of the point particle. However, because the gravitational and electromagnetic perturbations are coupled in such a spacetime, electromagnetic perturbations can be induced from gravitational ones. In this paper, we work within the Dudley–Finley approximation~\cite{Dudley:1977zz, Dudley:1978vd} (in which one of the two perturbative sectors is held fixed to restore separability of perturbation equations) and focus on the gravitational perturbation by ignoring its coupling to the electromagnetic one.
In this limit, the perturbation equations can be decomposed into Teukolsky-like, separable equations as in the Kerr case~\cite{Teukolsky:1973ha, Press:1973zz, Teukolsky:1974yv}.
Such a treatment has proven to be valid for the calculation of quasinormal mode frequencies when the charge is sufficiently small~\cite{Berti:2005eb,Saha:2025nsg}.
We will discuss the limitations of this approximation in Sec.~\ref{sec:discussion}.

In this section, we give an overview of the formalism for GW calculations.
We first present the explicit form of the Dudley--Finley equation and its formal solution using the Green's function in Secs.~\ref{sec:TSN} and \ref{sec:solution_TSN}.
In Sec.~\ref{sec:energy_flux}, we next derive the energy flux via GWs from a circular orbit of a test particle, which plays an important role in the evolution of asymmetric-mass mergers discussed in Sec.~\ref{sec:ITP_GWs}.

\subsection{The Dudley--Finley equation}
\label{sec:TSN}
Gravitational radiation at infinity is described by a Newman--Penrose curvature invariant $\psi_4 = -C_\mathrm{\alpha\beta\gamma\delta} n^\alpha \overline{m}^\beta n^\gamma \overline{m}^\delta$, where $C_\mathrm{\alpha\beta\gamma\delta}$ is the Weyl tensor, while $n^\alpha$ and $m^\alpha$ are the null tetrads defined by
\begin{align}
    n^\alpha &= \frac{1}{2\Sigma}\left( r^2+a^2, -\Delta, 0, a \right)\:, \\
    m^\alpha &= \frac{\overline{\rho}}{\sqrt{2}} \left( \mathrm{i}a\sin\theta, 0, 1, \frac{\mathrm{i}}{\sin\theta} \right)\:,
    \end{align}
with
\begin{equation}
    \rho(r, \theta) = \frac{1}{r-\mathrm{i}a\cos\theta}\:.
\end{equation}
An over bar represents the complex conjugate.
Ignoring the coupling of the gravitational perturbation to the electromagnetic one by setting the latter to zero, $\psi_4$ can be decomposed as
\begin{equation}
\label{eq:psi_4}
    \rho^{-4} \psi_4 = \sum_{l,m} \int \diff\omega\:\mathrm{e}^{-\mathrm{i}\omega t} R_{\ell m\omega}(r) \frac{{}_\mathrm{-2}S^{a\omega}_{\ell m}(\theta)}{\sqrt{2\pi}}\mathrm{e}^{\mathrm{i}m\phi}\:,
\end{equation}
where ${}_{-2}S^{a\omega}_{\ell m}$ represents the spin-weighted spheroidal harmonics.
$\psi_4$ is related to the strain at infinity by
\begin{equation}
\label{eq:h_psi_4}
    \psi_4(r\rightarrow\infty) = \frac{1}{2} \frac{\diff^2}{\diff t^2} \left(h_+ - \mathrm{i}h_\times\right)\:,
\end{equation}
where $h_{+, \times}$ are the plus and cross components.

Under the assumption that the coupling between gravitational and electromagnetic perturbations is ignored, the presence of $Q$ does not affect the angular structure of the Teukolsky equation from the Kerr case. Therefore, the angular sector is still governed by the spin-weighted spheroidal harmonics ${}_\mathrm{-2}S^{a\omega}_{\ell m}(\theta)$, which satisfy
\begin{equation}
\begin{split}
    \Bigg[ &\frac{1}{\sin\theta}\frac{\diff }{\diff \theta}\left( \sin^2\theta \frac{\diff}{\diff \theta} \right) - a^2\omega^2 \sin^2\theta - \frac{(m-2\cos\theta)}{\sin^2\theta}\\ &+ 4a\omega \cos\theta - 2 + 2ma\omega + \lambda_{\ell m \omega} \Bigg] {}_\mathrm{-2}S^{a\omega}_{\ell m}(\theta) = 0 \:,
\end{split}
\end{equation}
where $\lambda_{\ell m \omega}$ is the eigenvalue. We obtain the value of $\lambda_{\ell m \omega}$ using \texttt{qnm} package~\cite{Stein:2019mop} and fix the normalization so that
\begin{equation}
\label{eq:normalization_S_lm}
    \int^\pi_0 \big| {}_\mathrm{-2}S^{a\omega}_{\ell m}(\theta) \big|^2 \sin\theta \diff \theta=1\:.
\end{equation}

The radial function $R_{\ell m\omega}(r)$ satisfies the Dudley--Finley equation\footnote{In literature (e.g.~\cite{Berti:2005eb, Saha:2025nsg}), the Dudley--Finley equation refers to Eq.~\eqref{eq:radial_teukolsky} with $T_{\ell m n}=0$. In this work, we call Eq.~\eqref{eq:radial_teukolsky} the Dudley--Finley equation even though it has a non-vanishing source.
},
\begin{equation}
    \label{eq:radial_teukolsky}
    \Delta^2 \frac{\diff }{\diff r } \left( \frac{1}{\Delta} \frac{\diff R_{\ell m\omega}}{\diff r} \right) - V(r)R_{\ell m\omega} = T_{\ell m\omega}\:, 
\end{equation}
where $V(r)$ is given by 
\begin{equation}
    V(r) = -\frac{K^2 + 4\mathrm{i}(r-M)K}{\Delta} + 8\mathrm{i}\omega r + \lambda_{\ell m \omega}\:,
\end{equation}
with
\begin{equation}
    K(r) = (r^2+a^2)\omega - am\:.
\end{equation}
Reference~\cite{Geroch:1973am} gives the form of $T_{\ell m\omega}$,
\begin{equation}
\label{eq:T_lmw}
    T_{\ell m\omega} = 4\int \diff\Omega \diff t~\rho^{-5} \overline{\rho}^{-1} (B'_2+B'^*_2) \frac{{}_\mathrm{-2}S^{a\omega}_{\ell m}(\theta)}{\sqrt{2\pi}}\mathrm{e}^{-\mathrm{i}\omega t + \mathrm{i}m\phi}\:,
\end{equation}
where $B'_2$ and $B'^\ast_2$ are shown in Appendix~\ref{app:source_term}.
We will input the trajectories derived in Sec.~\ref{sec:stitch_ITP} and compute gravitational waveforms in Sec.~\ref{sec:GWs}.

\subsection{Solution to the Dudley--Finley equation}
\label{sec:solution_TSN}
We can formally solve Eq.~\eqref{eq:radial_teukolsky} with the Green's function method. To do so, we prepare two linearly independent homogeneous solutions, $R^\mathrm{in}_{\ell m\omega}$ and $R^\mathrm{up}_{\ell m\omega}$, whose asymptotic behaviors near the (outer) event horizon and infinity are given by
\begin{equation}
\label{eq:in_T}
R^\mathrm{in}_{\ell m\omega}( r_\ast) \rightarrow \left\{
\begin{array}{ll}
B^\mathrm{trans}_{\ell m\omega}\Delta^2\mathrm{e}^{-\mathrm{i}kr_\ast}\:, & r_\ast \rightarrow  -\infty\\
B^\mathrm{ref}_{\ell m\omega}r^3\mathrm{e}^{\mathrm{i}\omega r_\ast} + \frac{B^\mathrm{inc}_{\ell m\omega}}{r}\mathrm{e}^{-\mathrm{i}\omega r_\ast}\:, & r_\ast\rightarrow +\infty \\
\end{array}\;,
\right.
\end{equation}
\begin{equation}
\label{eq:up_T}
R^\mathrm{up}_{\ell m\omega}( r_\ast) \rightarrow \left\{
\begin{array}{ll}
D^\mathrm{ref}_{\ell m\omega}\Delta^2\mathrm{e}^{-\mathrm{i}kr_\ast} + D^\mathrm{inc}_{\ell m\omega}\mathrm{e}^{\mathrm{i}kr_\ast}\:, & r_\ast \rightarrow  -\infty\\
D^\mathrm{trans}_{\ell m\omega}r^3\mathrm{e}^{\mathrm{i}\omega r_\ast}\:, & r_\ast\rightarrow +\infty \\
\end{array}\;.
\right.
\end{equation}
Here $r_\ast$ is the tortoise coordinate defined by
\begin{equation}
    r_\ast = r + \frac{2M}{r_+-r_-}\left[ r_+\ln\left( \frac{r-r_+}{2M} \right) - r_-\ln\left( \frac{r-r_-}{2M} \right) \right]\:,
\end{equation}
where $k=\omega - ma/(r_+^2+a^2)$.
The $B_{\ell m\omega}$ and $D_{\ell m\omega}$ coefficients encode the amplitudes of the ingoing, outgoing, and transmitted wave components of the homogeneous solutions at the horizon and at infinity.  
The ``inc'', ``ref'', and ``trans'' subscripts indicate the incident, reflected, and transmitted parts of the radial wave, respectively.

Having such homogeneous solutions at hand, we can construct the Green's function so that the physical boundary conditions (no outgoing wave at the horizon and no ingoing wave at infinity) are satisfied:
\begin{equation}
\begin{split}
    \mathcal{G}_{\ell m \omega}(r, r') = \frac{1}{W_{\ell m\omega}} \Big[&R^\mathrm{up}_{\ell m \omega}(r) R^\mathrm{in}_{\ell m \omega}(r')\Theta(r-r')\\ &+ R^\mathrm{in}_{\ell m \omega}(r) R^\mathrm{up}_{\ell m \omega}(r')\Theta(r'-r) \Big]\:,
\end{split}
\end{equation}
Here $\Theta$ is the Heaviside's step function, and $W_{\ell m\omega}$ is the Wronskian given by
\begin{equation}
    W_{\ell m\omega} = 2\mathrm{i}\omega B^\mathrm{inc}_{\ell m\omega}D^\mathrm{trans}_{\ell m\omega}\:.
\end{equation}
The solution is given by the convolution integral of $T_{\ell m \omega}$ as
\begin{equation}
\label{eq:sol_T}
\begin{split}
    R_{\ell m \omega}(r) &= \int^{\infty}_{r_+} \diff r' \frac{\mathcal{G}_{\ell m \omega}(r, r')T_{\ell m \omega}(r')}{\Delta^2(r')} \\
    &= Z^\mathrm{H}_{\ell m \omega}(r) R^\mathrm{up}_{\ell m \omega}(r) + Z^\infty_{\ell m \omega}(r) R^\mathrm{in}_{\ell m \omega}(r)\:,
\end{split}
\end{equation}
where
\begin{align}
    Z^\mathrm{H}_{\ell m \omega}(r)  &= \frac{1}{ W_{\ell m \omega} } \int^{r}_{r_+} \diff r' \frac{R^\mathrm{in}_{\ell m \omega}(r')T_{\ell m \omega}(r')}{\Delta^2(r')}\:,\\
    Z^\infty_{\ell m \omega}(r) &= \frac{1}{W_{\ell m \omega}} \int^{\infty}_{r} \diff r' \frac{R^\mathrm{up}_{\ell m \omega}(r')T_{\ell m\omega}(r')}{\Delta^2(r')}\:.
\end{align}
In particular, the asymptotic behavior at infinity is
\begin{equation}
\label{eq:R_infty}
\begin{split}
    R_{\ell m\omega}(r\rightarrow \infty) &\rightarrow \frac{r^3 \mathrm{e}^{\mathrm{i}\omega r_\ast}}{2\mathrm{i} \omega B^\mathrm{inc}_{\ell m \omega}} \int^{\infty}_{r_+} \diff r' \frac{R^\mathrm{in}_{\ell m \omega}(r')T_{\ell m \omega}(r')}{\Delta^2(r')}\\
    &=: \tilde{Z}^\mathrm{H}_{\ell m\omega} r^3 \mathrm{e}^{\mathrm{i}\omega r_\ast}\:.
\end{split}
\end{equation}
We present the form of $\tilde{Z}^\mathrm{H}_{\ell m\omega}$ that is convenient for numerical calculations in Appendix~\ref{app:source_term}.
The gravitational waveform observed at a distance $D$ from the source is given by
\begin{equation}
    h_+ - \mathrm{i}h_\times = -\frac{2}{D} \sum_{\ell, m} \int^\infty_{-\infty} \diff \omega~\mathrm{e}^{-\mathrm{i}\omega t+\mathrm{i}m\phi} \frac{\tilde{Z}^\mathrm{H}_{\ell m\omega}}{\omega^2} \frac{{}_\mathrm{-2}S^{a\omega}_{\ell m}(\theta)}{\sqrt{2\pi}}\:.
\end{equation}

To derive $R^\mathrm{in,up}_{\ell m\omega}(r)$ and their coefficients, we first solve the Sasaki--Nakamura (SN) equation~\cite{Sasaki:1981sx} and convert the solution to $R_{\ell m \omega}$.
This is because, in contrast to Eq.~\eqref{eq:radial_teukolsky}, the SN equation has a short-range potential, allowing us to read off the coefficients without the divergent nature of $R_{\ell m\omega}$ at infinity (see Refs.~\cite{Lo:2023fvv, Yin:2025kls} for a review of SN formalism). 
The details of the SN equation of the Kerr--Newman BH and the conversion of the SN variable to $R_{\ell m\omega}$ are shown in Appendix~\ref{app:SN_eq}.

\subsection{GW energy flux from a circular orbit}
\label{sec:energy_flux}
Here, we evaluate the GW energy flux from a circular orbit on the equatorial plane. In this case, $Z^\mathrm{H}_{\ell m\omega}$ and $Z^\infty_{\ell m\omega}$ take non-zero values only at $\omega = \omega_m :=  m \Omega^\mathrm{(circ)}(r_\mathrm{orb})$, where $r_\mathrm{orb}$ is a radius of a circular orbit.

Given Eq.~\eqref{eq:h_psi_4} and the Isaacson stress-energy tensor~\cite{Isaacson:1968zza}, we can derive the GW energy flux at infinity as
\begin{equation}
\label{eq:flux_infty}
    \left( \frac{\diff E}{\diff t} \right)^\infty_\mathrm{GW} = \sum_{\ell,m} \frac{|Z^\mathrm{H}_{\ell m\omega_m}(r_\mathrm{orb})|^2}{4\pi \omega^2_m} =: \eta^2 \mathcal{F}^\infty(r_\mathrm{orb})\:,
\end{equation}
where we define $\mathcal{F}^\infty$ as the flux normalized by the mass ratio.

The flux at the horizon is derived utilizing the Teukolsky-Starobinsky identity~\cite{Teukolsky:1974yv} for $\psi_4$ and $\psi_0$, where the latter is the Weyl scalar containing ingoing gravitational radiation to the horizon.
The form is given by
\begin{equation}
\label{eq:flux_horizon}
    \left( \frac{\diff E}{\diff t} \right)^\mathrm{H}_\mathrm{GW} = \sum_{\ell,m} \alpha_{\ell m \omega_m} \frac{|Z^\infty_{\ell m\omega_m}(r_\mathrm{orb})|^2}{4\pi \omega^2_m} =: \eta^2 \mathcal{F}^\mathrm{H}(r_\mathrm{orb})\:.
\end{equation}
Similar to $\mathcal{F}^\infty$, $\mathcal{F}^\mathrm{H}$ is the normalized energy flux down to the horizon.
The coefficient $\alpha_{\ell m\omega}$, whose detailed derivation for the Kerr--Newman case is presented in Appendix~\ref{app:TS_identity}, is 
\begin{equation}
    \alpha_{\ell m\omega} = \frac{256(2Mr_+ - Q^2)^5 k (k^2+4\epsilon_\mathrm{KN}^2)(k^2+16\epsilon_\mathrm{KN}^2)\omega^3}{|C_{\ell m \omega}|^2}\:,
\end{equation}
with
\begin{align}
\epsilon_\mathrm{KN} =& \frac{\sqrt{M^{2} - \left(a^{2} + Q^{2}\right)}}{2\left(r_{+}^{2} + a^{2}\right)} \label{eq:eps} \\[2mm]
|C_{\ell m\omega}|^{2} =& 
\left[ \left(\lambda_{\ell m\omega} + 2\right)^{2} + 4 m a \omega - 4 a^{2} \omega^{2} \right] \nonumber \\
& \times 
\left( \lambda_{\ell m\omega}^{2} + 36 m a \omega - 36 a^{2} \omega^{2} \right) \notag \\ 
&+ \left( 2\lambda_{\ell m\omega} + 3 \right)\left( 96 a^{2} \omega^{2} - 48 m a \omega \right) 
\nonumber \\
&+ 144\,\omega^{2}\left(M^{2} - a^{2}\right)\:. \label{eq:C_abs}
\end{align}
Note that $|C_{\ell m\omega}|^2$ originates from the identity for the angular sector~\cite{Teukolsky:1974yv}, meaning that $Q$ does not appear in this case. 
On the other hand, the numerator of $\alpha_{\ell m\omega}$ is derived using the asymptotic expression of $\psi$ at the horizon, whose ingoing component is proportional to $\Delta^2$. 
It thus includes $Q$ through $\epsilon$ and $r_+$ (see Appendix~\ref{app:TS_identity} for more details).

\begin{figure}
    \centering
    \includegraphics[width=1.\linewidth]{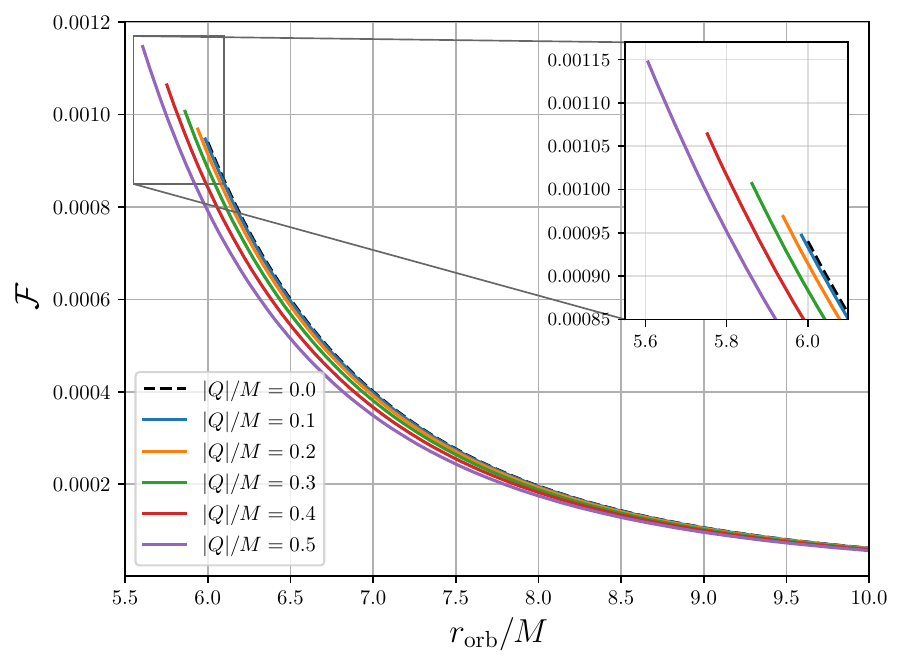}
    \caption{ Energy flux $\mathcal{F}$ for $|Q|/M=0, 0.1, 0.2, 0.3, 0.4,$ and $0.5$ with $a/M=0$.
    For illustration, we consider relatively large values of $|Q|$ here. The inset zoom in to the region near ISCO for each $Q/M$.}
    \label{fig:energy_flux_a_0}
\end{figure}

The total energy flux carried by GWs from a circular orbit at $r=r_\mathrm{orb}$ is given by
\begin{equation}
\label{eq:GW_rad} 
\begin{split}
    \left( \frac{\diff E}{\diff t} \right)_\mathrm{GW} &= \left( \frac{\diff E}{\diff t} \right)^\infty_\mathrm{GW} + \left( \frac{\diff E}{\diff t} \right)^\mathrm{H}_\mathrm{GW}\\
    &=: \eta^2 \mathcal{F}(r_\mathrm{orb})\:,
\end{split}
\end{equation}
where $\mathcal{F}:=\mathcal{F}^{\infty}+\mathcal{F}^\mathrm{H}$ is the normalized total energy flux. 
We include modes up to $\ell_{\max}=20,~30,$ and $60$ for $a/M=0.0,~0.5,$ and $0.9$, respectively, in Eqs.~\eqref{eq:flux_infty} and \eqref{eq:flux_horizon}. 
For $a/M=0$ and $0.5$, these choices suppress the mode–truncation error in $\mathcal{F}$ to the $\mathcal{O}(10^{-8})$ level (see Fig.~2 of Ref.~\cite{Khalvati:2025znb}), while for $a/M=0.9$, we follow Ref.~\cite{Khalvati:2025znb} and treat $\ell_{\max}=60$ as sufficiently
converged in the rapidly spinning regime.\footnote{The required accuracy is fixed separately for each BH spin so that the threshold value of $Q$ that produces a dephasing of one radian in EMRIs (Sec.~\ref{sec:EMRI}) corresponds to flux differences at the $\sim10^{-7}$ level, which remain safely larger than the residual mode–truncation error even with the cutoff at finite $\ell_{\max}$.}
Figure~\ref{fig:energy_flux_a_0} shows the normalized total energy flux $\mathcal{F}$ for $|Q|/M=0,~0.1,~0.2,~0.3,~0.4,$ and $0.5$ with $a/M=0$. 
Each curve starts from $r_\mathrm{orb}/M=10$ to $r_\mathrm{ISCO}$.
This figure indicates that, for a fixed $r_\mathrm{orb}$, the value of $\mathcal{F}$ becomes smaller as $|Q|/M$ is larger, leading to a slower orbital-radius evolution.

To check the validity of our numerical code, we compare the results of the flux using our code and \texttt{Black hole perturbation toolkit}~\cite{BHPToolkit} in Appendix~\ref{app:code_check}. 
We find that the relative error is of order  $10^{-2}~\%$ in the Schwarzschild case and of $10^{-1}~\%$ in the Kerr case. 

\section{Inspiral-transition-plunge gravitational waveforms}
\label{sec:ITP_GWs}
Our purpose is to compute, for the first time, inspiral-transition-plunge waveforms from asymmetric-mass binaries with the beyond-Kerr framework, taking Kerr-Newman as an example.
To achieve this, we need the worldline of the infalling secondary that we input in Eq.~\eqref{eq:T_lmw}. 

The process of an asymmetric merger is typically divided into three stages: the \ins, \tra, and \plu. We combine these three phases closely following Apte and Hughes 2019~\cite{Apte:2019txp} (hereafter, AH19). 
In Sec.~\ref{sec:ITP}, we give an overview of the \ins, \tra, and \plu, and a qualitative treatment for each regime. 
We then stitch the trajectories of the secondary object in these three regimes based on AH19 in Sec.~\ref{sec:stitch_ITP}. 
Finally, we show the resulting waveforms in Sec.~\ref{sec:ITP_GWs}.

\subsection{Adiabatic inspiral, transition, and plunge}
\label{sec:ITP}
During the inspiral phase, the secondary object evolves adiabatically due to GW radiation. 
The orbital radius changes much more slowly compared to the orbital period. 
Therefore, the motion can be approximated as a sequence of quasi-circular orbits characterized by the energy $\Ecirc(r_\mathrm{orb})$ and angular momentum $\Lcirc(r_\mathrm{orb})$ at each orbital radius $r_\mathrm{orb}$.

As the secondary approaches the ISCO, the adiabatic approximation gradually breaks down. The radial motion begins to accelerate, influenced by both radiation reaction and the strong-field gravitational attraction. This intermediate stage is known as the transition regime.

In the final plunge phase, the radiation-reaction timescale becomes much shorter than the orbital timescale. The motion is well approximated by a geodesic, and the secondary plunges into the primary BH.

\subsubsection{\textbf{Inspiral}}
\label{sec:inspiral}

Up to the near-ISCO region, the secondary's orbit evolves adiabatically due to gravitational radiation. 
The trajectory can be regarded as a sequence of circular orbits at each orbital radius.  
For a given radius $r$, substituting the energy $\hat{E}^\mathrm{(circ)}(r)$ and angular momentum $\hat{L}^\mathrm{(circ)}(r)$ for a circular orbit into the geodesic equations, Eqs.~\eqref{eq:t_geo} and~\eqref{eq:phi_geo}, determines the corresponding circular orbit.  
Assuming that the only mechanism driving the radial evolution is the GW radiation reaction, the slow drift of the orbital radius is then governed by
\begin{equation}
\label{eq:drdt_ins}
    \frac{\diff r}{\diff \lambda}
    = \frac{\diff r}{\diff t}\,\frac{\diff t}{\diff \lambda}
    = -\,\frac{\left(\frac{\diff \hat{E}}{\diff t}\right)_\mathrm{GW}}
           {\frac{\diff \hat{E}^\mathrm{(circ)}}{\diff r}}
      \frac{\diff t}{\diff \lambda}
    \propto \eta \:,
\end{equation}
where $\diff t / \diff \lambda$ is given by Eq.~\eqref{eq:t_geo}.

\subsubsection{\textbf{Transition}}
\label{sec:transition}
As the secondary approaches the ISCO, the transition regime begins (e.g.~\cite{Ori:2000zn}). 
The basic idea to derive the equation of motion in the transition regime is to expand Eq.~\eqref{eq:r_geo} around the ISCO. 
Since $r_\mathrm{ISCO}$ satisfies $R = 0, \partial R/\partial r=0$, and $\partial^2 R/\partial r^2 =0$, we can expand $R(r, \hat{E}, \hat{L})$ around $r=r_\mathrm{ISCO}$ as
\begin{equation}
\label{eq:expand_R_ISCO}
\begin{split}
    R(r, \hat{E}, \hat{L}) = &~\frac{1}{6} \left(\frac{\partial^3 R}{\partial r^3}\right)_\mathrm{ISCO} x^3\\ &+ \left[ \left( \frac{\partial^2 R}{\partial r \partial \hat{E}}\right)_\mathrm{ISCO}\delta \hat{E} + \left( \frac{\partial^2 R}{\partial r \partial \hat{L}}\right)_\mathrm{ISCO}\delta \hat{L} \right] x\:,
\end{split}
\end{equation}
with $x:=r-r_\mathrm{ISCO}$, $\delta \hat{E} := \hat{E}-\hat{E}_\mathrm{ISCO}$, and $\delta \hat{L} := \hat{L}-\hat{L}_\mathrm{ISCO}$. 
Here, we ignore terms that do not depend on $x$. 
Following Ref.~\cite{Ori:2000zn}, we assume that the flux in the transition regime is the same as that at ISCO, and that the orbit is nearly circular. 
Then, $\hat{E}$ and $\hat{L}$ can be approximated by 
\begin{align}
    \hat{E} &= \hat{E}_\mathrm{ISCO}+\left( \frac{\diff \hat{E}}{\diff \lambda} \right)_\mathrm{ISCO} (\lambda -\lambda_\mathrm{ISCO}) \:,\label{eq:delta_E_tra}\\
    \hat{L} &= \hat{L}_\mathrm{ISCO}+\left( \frac{\diff \hat{L}}{\diff \lambda} \right)_\mathrm{ISCO} (\lambda -\lambda_\mathrm{ISCO}) \:\label{eq:delta_L_tra},
\end{align}
with 
\begin{align}
    \left( \frac{\diff \hat{E}}{\diff \lambda} \right)_\mathrm{ISCO} &:= -\eta~\kappa_\mathrm{ISCO}\:, \\
    \left( \frac{\diff \hat{L}}{\diff \lambda} \right)_\mathrm{ISCO} &:= -\eta~\Omega_\mathrm{ISCO}^{-1}~\kappa_\mathrm{ISCO}\:.
\end{align}
$\lambda_\mathrm{ISCO}$ is the Mino time at which $\hat{L}=\hat{L}_\mathrm{ISCO}$, and $\kappa_\mathrm{ISCO}$ is the GW energy flux at ISCO in terms of $\lambda$ defined by $\kappa_\mathrm{ISCO}:=\mathcal{F}_\mathrm{ISCO}  \cdot \diff t/\diff \lambda (r_\mathrm{ISCO}, \hat{E}_\mathrm{ISCO}, \hat{L}_\mathrm{ISCO})$.

Taking a derivative of Eq.~\eqref{eq:r_geo} with respective to $r$, we can derive equation of motion in the transition regime as
\begin{equation}
\label{eq:EOM_tra}
    \frac{\diff^2 x}{\diff \lambda^2} = -Ax^2 - \eta B (\lambda-\lambda_\mathrm{ISCO})\:,
\end{equation}
where
\begin{align}
    A &:= -\frac{1}{4} \left(\frac{\partial^3 R}{\partial r^3}\right)_\mathrm{ISCO}\:,\\
    B &:= -\frac{1}{2} \left[ \left( \frac{\partial^2 R}{\partial r \partial \hat{E}}\right)_\mathrm{ISCO} + \Omega_\mathrm{ISCO}^{-1}\left( \frac{\partial^2 R}{\partial r \partial \hat{L}}\right)_\mathrm{ISCO} \right]\kappa_\mathrm{ISCO}\:.
\end{align} 
The first term in Eq.~\eqref{eq:EOM_tra} represents the plunge, while the second encodes the effect of radiation reaction.
We can convert Eq.~\eqref{eq:EOM_tra} into a dimensionless form,
\begin{equation}
\label{eq:dimless_EOM_tra}
    \frac{\diff^2 X }{\diff L^2} = -X^2 - L\:,
\end{equation}
by scaling variables with the following relations:
\begin{align}
    x/M &:= \eta^{2/5} B^{2/5} A^{-3/5} X \label{eq:scaling_X} \:,\\
    M(\lambda-\lambda_\mathrm{ISCO}) &:= \eta^{-1/5} (AB)^{-1/5} L\:. \label{eq:scaling_lambda}
\end{align}
We expect the solution $X$ to smoothly connect to the adiabatic inspiral, indicating that 
\begin{equation}
\label{eq:sol_X_negative_T}
    X = \sqrt{-L}\:,
\end{equation}
for large negative $L$. 
We can numerically derive the solution $X$ using Eq.~\eqref{eq:sol_X_negative_T} as an initial condition.
The solution is universal in that, once obtained, it allows us to derive $r$ and $\lambda$ through Eqs.~\eqref{eq:scaling_X} and \eqref{eq:scaling_lambda} for any $a$ and $\eta$, as long as $\eta$ is sufficiently small.

Importantly, the derivation of Eq.~\eqref{eq:EOM_tra} does not rely on any assumptions unique to the Kerr BH.
The key requirement is that the radial motion near the ISCO is assumed to be described by the leading terms in the expansion of $R$. Therefore, the framework can be straightforwardly applied to the Kerr--Newman BH.

\subsubsection{\textbf{Plunge}}
\label{sec:plunge}
As the radiation-reaction timescale becomes much shorter than the orbital timescale, the plunge trajectory is well described by a geodesic motion (Eqs.~\eqref{eq:t_geo}--\eqref{eq:phi_geo}) with conserved energy and angular momentum at the plunge stage,
\begin{align}
    \hat{E}_\mathrm{plunge} &= \hat{E}_\mathrm{ISCO} - \delta \hat{E}_\mathrm{trans}\:,\\
    \hat{L}_\mathrm{plunge} &= \hat{L}_\mathrm{ISCO} - \delta \hat{L}_\mathrm{trans}\:,
\end{align}
where $\delta \hat{E}_\mathrm{trans}$ and $\delta \hat{L}_\mathrm{trans}$.
The values of $\delta \hat{E}_\mathrm{trans}$ and $\hat{L}_\mathrm{trans}$ depend on at which $L$ the transition begins and ends, which will be discussed in the next subsection.

\subsection{Construction of the inspiral-transition-plunge worldline}
\label{sec:stitch_ITP}

We now connect the inspiral, transition, and plunge regimes into a continuous worldline. 
Following AH19, we first ensure the smoothness of energy and angular momentum across the inspiral–transition boundary in Sec.~\ref{sec:continuity_energy_angular_momentum}, and then determine the time range for the transition regime, $L_\mathrm{i} \leq L \leq L_\mathrm{f}$, based on the criteria introduced in Sec.~\ref{sec:stitch_criteria}.

\subsubsection{\textbf{Ensuring continuity of energy and angular momentum}}
\label{sec:continuity_energy_angular_momentum}

AH19 pointed out that directly using the expressions for energy and angular momentum in the transition regime, Eqs.~\eqref{eq:delta_E_tra} and \eqref{eq:delta_L_tra}, leads to a discontinuity at the inspiral–transition junction, that arises because $\hat{E}$ and $\hat{L}$ do not evolve linearly in $\lambda$. 
This discontinuity causes unphysical artifacts in the worldline through the transformation of $\lambda$ into the coordinate time $t$ (see Fig.~3 in AH19).

To resolve this issue, we use a model for $\hat{E}$ and $\hat{L}$ proposed in AH19, which includes phenomenological corrections up to cubic order in $\lambda$. For example, $\hat E$ is given by
\begin{equation}
\label{eq:E_tra_refined}
    \hat{E} = \hat{E}_\mathrm{ISCO} + \lambda \left( \frac{\diff \hat{E}}{\diff \lambda} \right)_\mathrm{ISCO} + \frac{\lambda^2}{2} \mathcal{E}_2 + \frac{\lambda^3}{6} \mathcal{E}_3\:,
\end{equation}
where $\lambda=0$ is defined such that $\hat{E}(\lambda=0)=\hat{E}_\mathrm{ISCO}$ and $\diff \hat{E}/\diff \lambda(\lambda=0) = (\diff \hat{E}/\diff \lambda)_\mathrm{ISCO}$. The coefficients $\mathcal{E}_{2,3}$ are determined to smoothly match the energy description in the inspiral up to quadratic order in $\lambda$ at $\lambda = \lambda_\mathrm{i}$:
\begin{widetext}
\begin{align}
    \mathcal{E}_2 &= \frac{2}{\lambda^2_\mathrm{i}} \left\{ 3\hat{E}(\lambda_\mathrm{i})-3\hat{E}_\mathrm{ISCO} - \lambda_\mathrm{i}\left[ 2\left( \frac{\diff \hat{E}}{\diff \lambda} \right)_\mathrm{ISCO} + \left( \frac{\diff \hat{E}}{\diff \lambda} \right)_{\lambda=\lambda_\mathrm{i}} \right] \right\}\:,\\
    \mathcal{E}_3 &= \frac{6}{\lambda^3_\mathrm{i}} \left\{ 2\hat{E}(\lambda_\mathrm{i})-2\hat{E}_\mathrm{ISCO} + \lambda_\mathrm{i}\left[ \left( \frac{\diff \hat{E}}{\diff \lambda} \right)_\mathrm{ISCO} + \left( \frac{\diff \hat{E}}{\diff \lambda} \right)_{\lambda=\lambda_\mathrm{i}} \right] \right\}\:.
\end{align}
\end{widetext}
An analogous expression is used for $\hat{L}$.

\subsubsection{\textbf{Criteria for choosing the start and end of the transition: $L_\mathrm{i}$ and $L_\mathrm{f}$}}
\label{sec:stitch_criteria}

We follow the same criteria as AH19 for determining the start and final time of the transition regime.
Their criteria are summarized as follows:
\begin{itemize}
    \item \textbf{Initial time $L_\mathrm{i}$}~:  
    AH19 recommends the range $-4 \leq L_\mathrm{i} \leq -1.4$ for the following reasons:
    \begin{itemize}
        \item[(a)] If $|L_\mathrm{i}|$ is too large, the refined energy expression~Eq.~\eqref{eq:E_tra_refined} and the BH perturbation expression~Eq.~\eqref{eq:delta_E_tra} no longer match.  
        In particular, at $L_\mathrm{i} = -4$, the fractional error between these two expressions is found to be about $5~\%$, which motivates taking $L_\mathrm{i} = -4$ as the lower bound.
        \item[(b)] If $|L_\mathrm{i}|$ is too small, post-adiabatic corrections become non-negligible, indicating that the adiabatic approximation has already begun to break down.  
        The first correction term in Eq.~\eqref{eq:EOM_tra} reaches $5~\%$ of the leading term around $L_\mathrm{i} = -1.4$, which sets the upper boundary of the allowed range.
    \end{itemize}
    \item \textbf{Final time $L_\mathrm{f}$}~:  
    AH19 recommends choosing $L_\mathrm{f}$ within the range $2.2 \leq L_\mathrm{f} \leq 2.5$, based on the behavior of Eq.~\eqref{eq:EOM_tra}.
    \begin{itemize}
        \item[(c)] If $L_\mathrm{f}$ is too small, the contribution of the non–free-fall term (the second term) in Eq.~\eqref{eq:EOM_tra} is not negligible, indicating that the motion has not yet reached the free-fall regime.  
        At $L_\mathrm{f} = 2.2$, this contribution falls below $5~\%$, marking a reasonable starting point for the plunge.
        \item[(d)] If $L_\mathrm{f}$ is too large, the higher-order correction neglected in Eq.~\eqref{eq:EOM_tra} becomes significant, and the truncated expansion ceases to be accurate.  
        Indeed, the leading correction exceeds $5~\%$ of the leading term at $L_\mathrm{f} = 2.5$, which therefore serves as a practical upper bound for the transition regime.
    \end{itemize}
\end{itemize}

AH19 mentions that, within these ranges, the resulting worldline is largely insensitive to the choice of $L_\mathrm{i}$, 
but can vary depending on the chosen value of $L_\mathrm{f}$. 
However, Ref.~\cite{Lim:2019xrb}, a follow-up work of AH19, demonstrates that the excitation of quasi-normal modes during the plunge is not significantly affected by this choice, implying that relevant physics is not sensitive to this choice.

Note that the conditions (b), (c), and (d) do not depend on the presence of $Q$.
This is because they are derived from Eq.~\eqref{eq:EOM_tra}, where we work with the normalized quantities $X$ and $L$, and the equation is independent of $Q$.
On the other hand, condition (a) depends on $Q$, but we have confirmed that it still holds for the values of $Q$ studied in this work. 
Given these, we fix $L_\mathrm{i} = -3$ and $L_\mathrm{f} = 2.35$, which both lie near the middle of these recommended ranges and are the values adopted in AH19.

\begin{figure} 
    \centering 
    \includegraphics[width=0.9\linewidth]{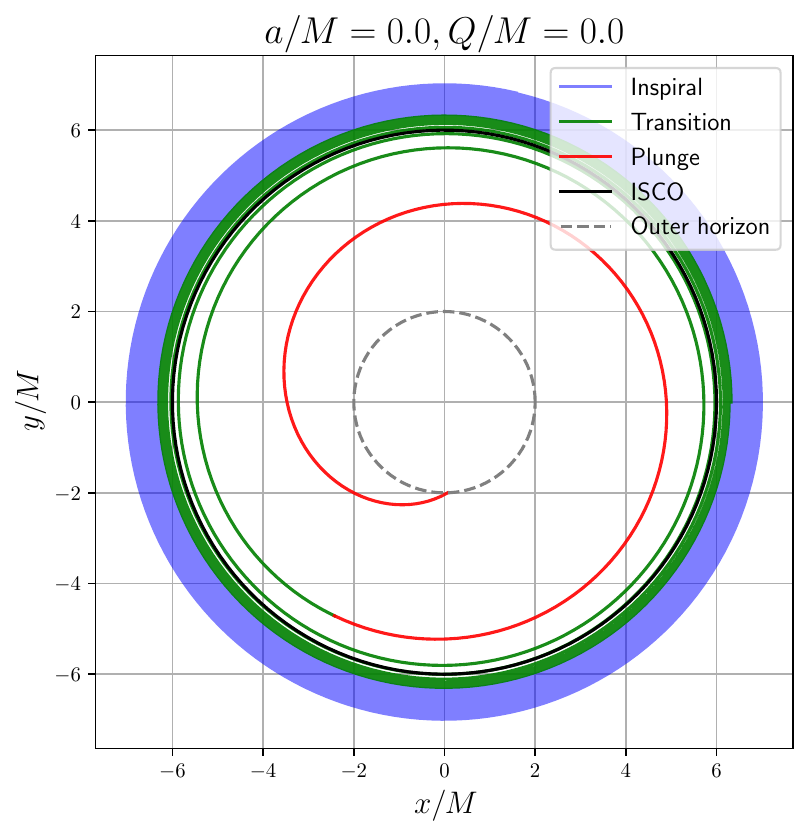} 
    \caption{Inspiral-transition-plunge trajectory on the equatorial plane for $a/M=0, Q/M=0,$ and $\eta=10^{-3}$, starting at $r=7M$.
    Blue, green, and red lines show the inspiral, transition, and plunge, respectively.
    Black solid and gray dashed lines depict the ISCO and outer horizon.} 
    \label{fig:ITP_trajectory_Schwarzschild} 
\end{figure} 

\begin{figure}
    \centering
    \includegraphics[width=\linewidth]{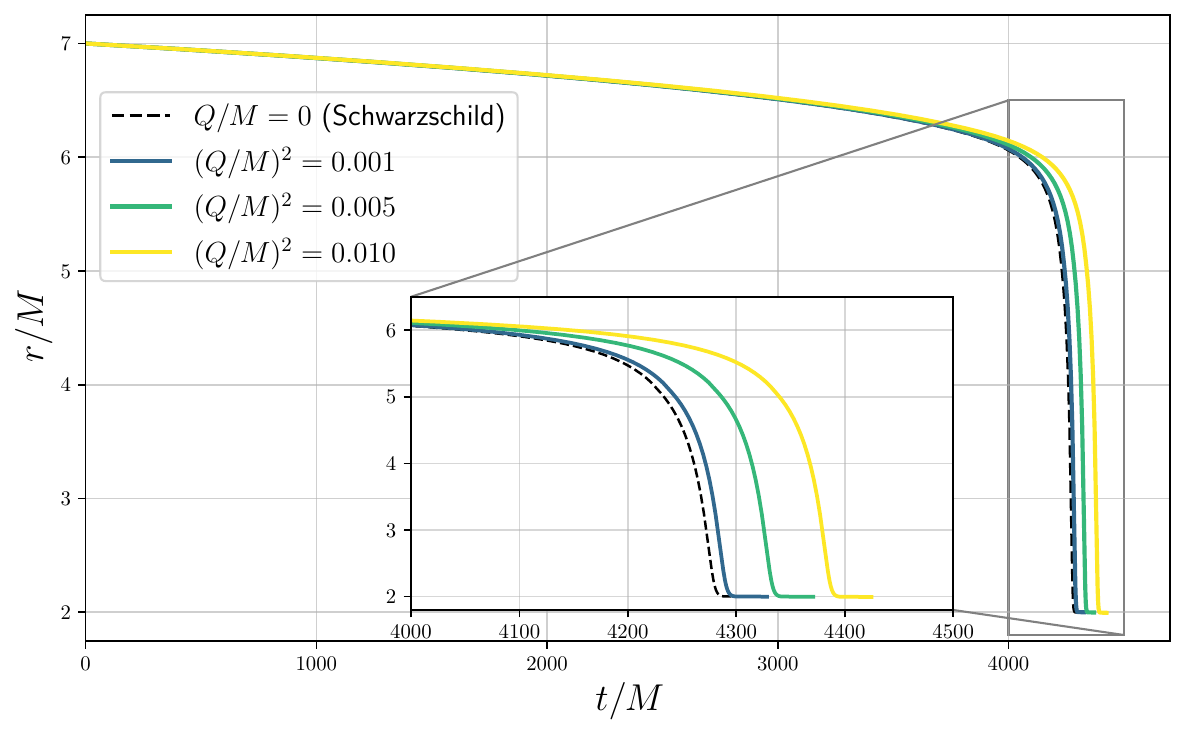}
    \caption{
    Time evolution of the orbital radius $r/M$ for inspiral–transition–plunge trajectories with different values of $Q/M$ and a fixed value of $a/M=0$.
    Each trajectory begins at $r = 7M$.
    The black dashed curve corresponds to the Schwarzschild case, while the colored solid curves show the charged cases with $(Q/M)^2 = 0.001$ (navy), $0.005$ (green), and $0.01$ (yellow).
    }
    \label{fig:t_vs_r_Q}
\end{figure}

Figure~\ref{fig:ITP_trajectory_Schwarzschild} shows the inspiral-transition-plunge trajectory on the equatorial plane for $a/M=0, Q/M=0,$ and $\eta=10^{-3}$ that is constructed based on the procedure outlined in Sec.~\ref{sec:stitch_ITP}.
Blue, green, and red lines depict the inspiral, transition, and plunge, respectively.
The black solid line marks the ISCO and outer horizon, while the gray dashed line marks the outer horizon.

\begin{figure*}
    \centering
    \includegraphics[width=1.\linewidth]{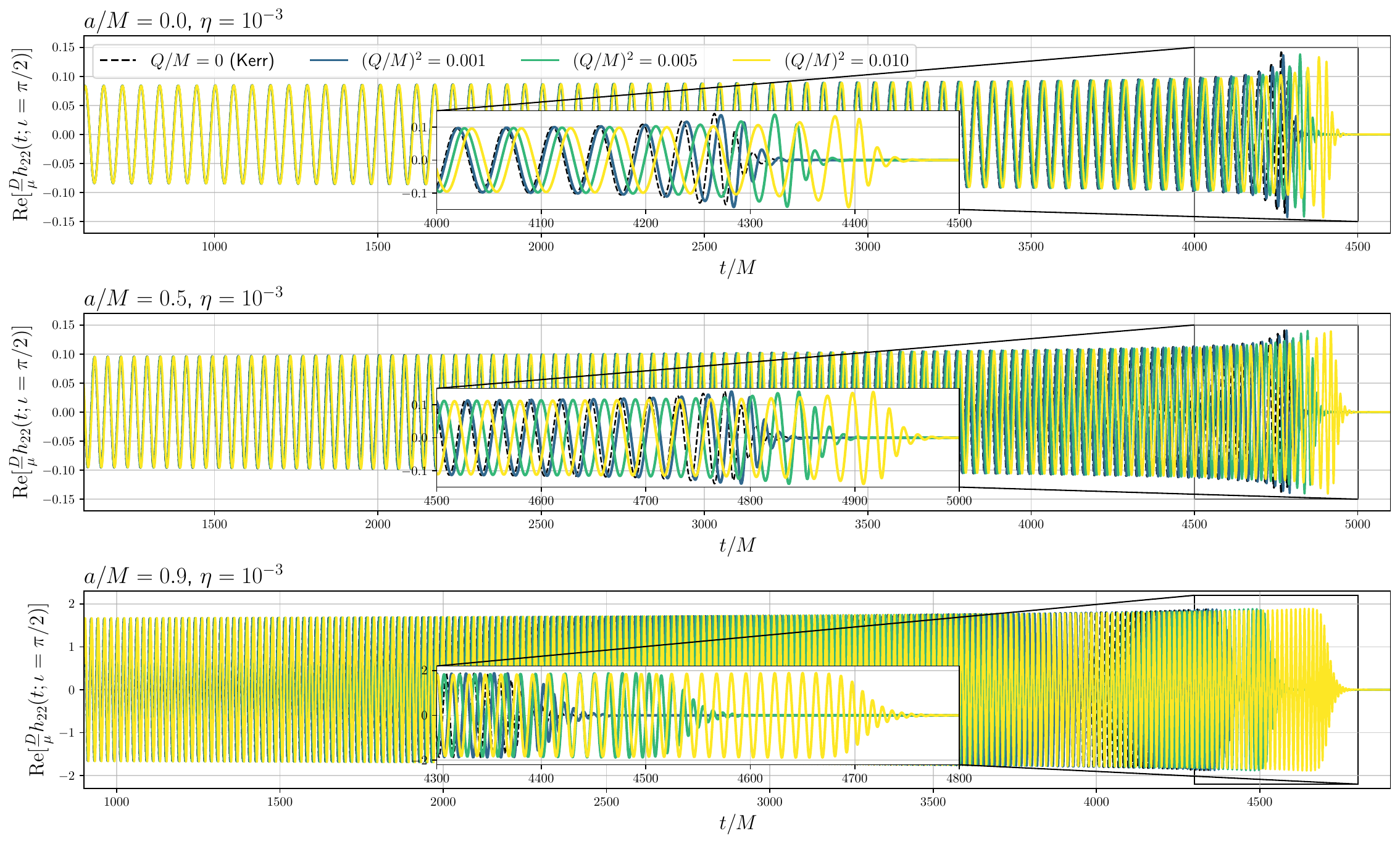}
    \caption{
    Gravitational waveforms of the $(\ell,m)=(2,2)$ mode are shown for intermediate–mass–ratio binary mergers with $\eta = 10^{-3}$, evaluated at an inclination angle $\iota = \pi/2$, for the BH spins of $a/M = 0$ (top), $0.5$ (middle), and $0.9$ (bottom), respectively. 
    The black dashed curve represents the Kerr case ($Q/M = 0$), while the colored solid curves correspond to $(Q/M)^2 = 0.001$ (navy), $0.005$ (green), and $0.01$ (yellow). 
    The orbital evolution starts at $r = 7M$, $5.5M$, and $3.2M$ for $a/M = 0$, $0.5$, and $0.9$, respectively. 
    Each panel displays the waveform over the final $4000M$ in time, with an inset focusing on the last $500M$.
    }
    \label{fig:GW}
\end{figure*}

Figure~\ref{fig:t_vs_r_Q} depicts the inspiral–transition–plunge trajectories for $a/M=0$ and different values of $Q/M$.
As $|Q|$ increases, the radiation flux at a given radius deceases (see Fig.~\ref{fig:energy_flux_a_0}), which slows down the orbital evolution and causes the trajectory to remain at larger radii for a longer time.
These trajectories are used in the top panel of Fig.~\ref{fig:GW}.

\subsection{Numerical waveforms}
\label{sec:GWs}
Throughout the trajectory, we use the source term presented in Appendix~\ref{app:source_term}.
Figure~\ref{fig:GW} presents the $(\ell,m)=(2,2)$ gravitational waveforms from intermediate–mass–ratio binary mergers with $\eta = 10^{-3}$, evaluated at an inclination angle of $\iota = \pi/2$\footnote{Because the spheroidal harmonic ${}_{-2}S^{a\omega}_{\ell m}(\iota)$ depends on $\omega$, the inverse Fourier transform must be performed while retaining this $\omega$–dependence as long as we work within this basis. 
Consequently, the resulting time-domain signal $h_{22}(t)$ inherits an explicit dependence on the inclination angle $\iota$.}. 
The top, middle, and bottom panels correspond to BH spins of $a/M = 0$, $0.5$, and $0.9$, respectively. 
For each spin, the black dashed curve shows the Kerr waveform ($Q/M = 0$), while the colored solid curves represent charged cases with $(Q/M)^2 = 0.001$, $0.005$, and $0.01$. 
Toward the ringdown phase, the accumulated phase offset among the waveforms becomes more evident, as the differences accumulated during the inspiral and transition are reflected in a visible shift of the late-time oscillations. 
This behavior becomes more pronounced for higher spins, where stronger frame dragging results in more orbital cycles, leading to a larger phase accumulation.

\section{Observational prospects}
\label{sec:observation_prospect}

In this section, we explore the observational prospects of the inspiral–transition–plunge waveforms constructed in the previous section, organizing the discussion around three guiding questions.  
Anticipating detections of intermediate–mass–ratio binary BH mergers with the Einstein Telescope (ET), we ask:
\begin{enumerate}
    \item In which region of parameter space does accurate modeling of the post-inspiral portion of the signal provide a significant advantage over the case with inspiral alone?
    \item How much improvement we gain on constraining the BH charge by adding the post-inspiral information on top of the inspiral one?

\end{enumerate}
In addition, looking ahead to the EMRIs observed by
space-based GW interferometers, such as LISA, we further ask:
\begin{enumerate}
\setcounter{enumi}{2}
    \item How small a charge-to-mass ratio can be distinguished through the accumulated dephasing in EMRIs?
\end{enumerate}

We address the first two questions in Secs.~\ref{sec:ET_SNR} and \ref{sec:distinguishability_IMR}, and turn to the EMRI dephasing analysis in Sec.~\ref{sec:EMRI} below.

\subsection{Inspiral and post-inspiral SNRs of intermediate-mass-ratio binary BH mergers with ET}
\label{sec:ET_SNR}

We evaluate the inspiral and post-inspiral SNRs of intermediate–mass-ratio binary mergers detectable by ET.
Our goal in this subsection is to identify the region of parameter space, specifically the primary mass and spin, where the inspiral and post-inspiral phases contribute comparably to the total SNR.
Such systems are of particular interest because they provide a unique opportunity to test the consistency between the inspiral and the post-inspiral data.

Since the presence of charge is not expected to significantly affect the overall SNR balance as long as $|Q|/M$ is small, we focus here on the Kerr cases. 
To explore the SNR ratio between the inspiral and post-inspiral regimes systematically, we employ the surrogate model \texttt{BHPTNRSur2dq1e3}~\cite{Rink:2024swg}, which provides approximations of the spherical-harmonic GW modes for mass ratios $10^{-4} \leq \eta \leq 1/3$ and primary spins $-0.8 \leq a/M \leq 0.8$, assuming a nonspinning secondary.
In the small-mass-ratio regime, the model is trained on waveforms computed within the point-particle BH perturbation framework, the same as the approach adopted in the previous sections.
We use this surrogate model instead of our inspiral-transition-plunge waveforms because it enables us to densely sample parameter space more efficiently.

Throughout this analysis, we fix the mass ratio and luminosity distance to $\eta = 10^{-2}$ and $D = 5~\mathrm{Gpc}$, corresponding to a redshift $z = 0.78$ under the $\Lambda$-cold-dark-matter cosmology model with parameters reported in Ref.~\cite{Planck:2015fie}.
For the ET configuration, we assume the ET-D sensitivity and adopt a triangular design consisting of three interferometers with $\pi/3$ opening angles~\cite{Hild:2010id}.

\subsubsection{\textbf{Calculation of SNR}}
We denote the Fourier transform of the detector output by
\begin{equation}
    \tilde{h}(f) = F_+ \tilde{h}_+(f) + F_\times \tilde{h}_\times(f)\:,
\end{equation}
where $F_{+,\times}$ are the antenna-pattern functions. 
As our focus is on understanding the values of the primary mass in source frame $M$ and spin $a$ suitable for the IMR-consistency type test of GR with detections by ET, we average over the sky location, polarization, and inclination angles of the source. 
We here focus on the dominant $(\ell,m)=(2,2)$ mode.
The averaged response of a single detector of ET, whose individual Michelson interferometers have an opening angle of $\gamma = \pi/3$ corresponding to a triangle-shaped design~\cite{Babak:2021mhe}, 
\begin{equation}
\tilde{h}_\mathrm{ave}(f)=\frac{2}{5}\sin{\gamma}~ \tilde{h}_{22}(f)=\frac{\sqrt{3}}{5} \tilde{h}_{22}(f)\:.
\end{equation}

We define the noise-weighted inner product between two waveforms $h_1$ and $h_2$
over a frequency interval $f_1 \le f \le f_2$ as
\begin{equation}
    \langle h_1 | h_2 \rangle (f_1, f_2)
    := 4 \, \mathrm{Re}
    \left[
        \int_{f_1}^{f_2}
        \frac{\tilde{h}_1(f)\tilde{h}^{\ast}_2(f)}{S_\mathrm{n}(f)}
        \, \diff f
    \right],
\end{equation}
where ${}^{\ast}$ denotes complex conjugation and $S_\mathrm{n}(f)$ is the designed noise power spectral density (PSD) of a single detector.
Using this definition, we introduce the inspiral (ins) and post-inspiral (pins) single-detector SNRs as
Here, $f_\mathrm{min}$ is set to $1~\mathrm{Hz}$, and $f_\mathrm{max}$ is chosen such that the integral converges depending on the total mass (typically $\sim 1000~\mathrm{Hz}$). 
The $(2,2)$ mode frequency at ISCO of the primary BH is defined as $f_\mathrm{ISCO}:=2\Omega_\mathrm{ISCO}/(2\pi(1+z))$. 
Since ET will consist of three interferometers rotated by $\pi/3$ with respect to each other, corresponding network SNRs are given by 
$\rho_\mathrm{ins} := \sqrt{3}\,\rho^\mathrm{single}_\mathrm{ins}$ and 
$\rho_\mathrm{pins} := \sqrt{3}\,\rho^\mathrm{single}_\mathrm{pins}$.

We compute $\rho_\mathrm{ins}$ and $\rho_\mathrm{pins}$ by varying the primary mass in the source frame $M$ and the primary spin $a/M$ over the ranges $10^2~M_\odot \leq M \leq 10^4~M_\odot$ and $0 \leq a/M \leq 0.9$\footnote{Note that we use \texttt{BHPTNRSur2dq1e3} as an extrapolation in parameter space where $a/M$ is greater than $0.8$.}, for a fixed mass ratio $\eta = 10^{-2}$ and a luminosity distance $D = 5~\mathrm{Gpc}$ ($z = 0.78$).

\subsubsection{\textbf{Results}}
\begin{figure}
    \centering
    \includegraphics[width=1.\linewidth]{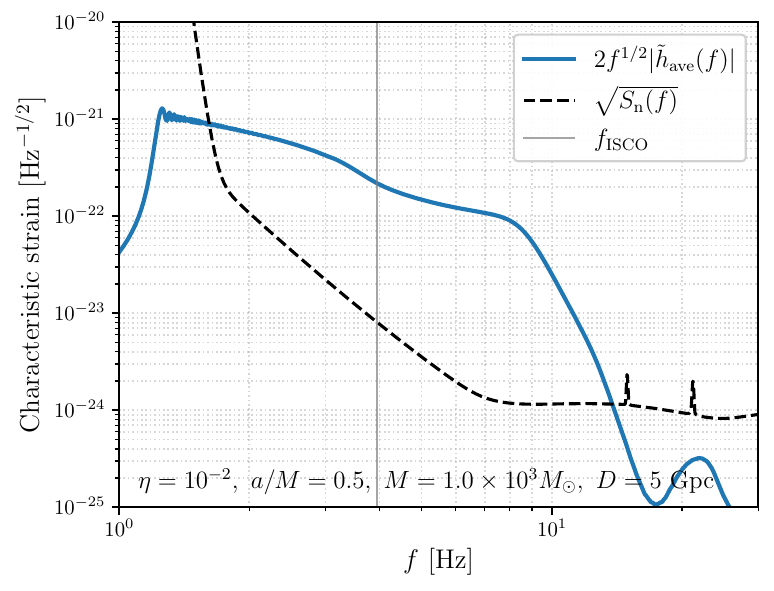}
    \caption{
    The angular-averaged frequency-domain strain $2f^{1/2}|\tilde{h}_\mathrm{ave}(f)|$ of an intermediate-mass-ratio binary BH merger 
    with $\eta=10^{-2}$, $a/M=0.5$, $M=10^3\,M_\odot$, and luminosity distance $D=5~\mathrm{Gpc}$ (blue), 
    compared with the square root of the designed noise PSD (black dashed). 
    The vertical gray line marks the ISCO frequency of the $(2,2)$ mode $f_\mathrm{ISCO}$.
    The strain is derived using \texttt{BHPTNRSur2dq1e3}.
    }
    \label{fig:strain_plot}
\end{figure}

\begin{figure}
    \centering
    \includegraphics[width=1.\linewidth]{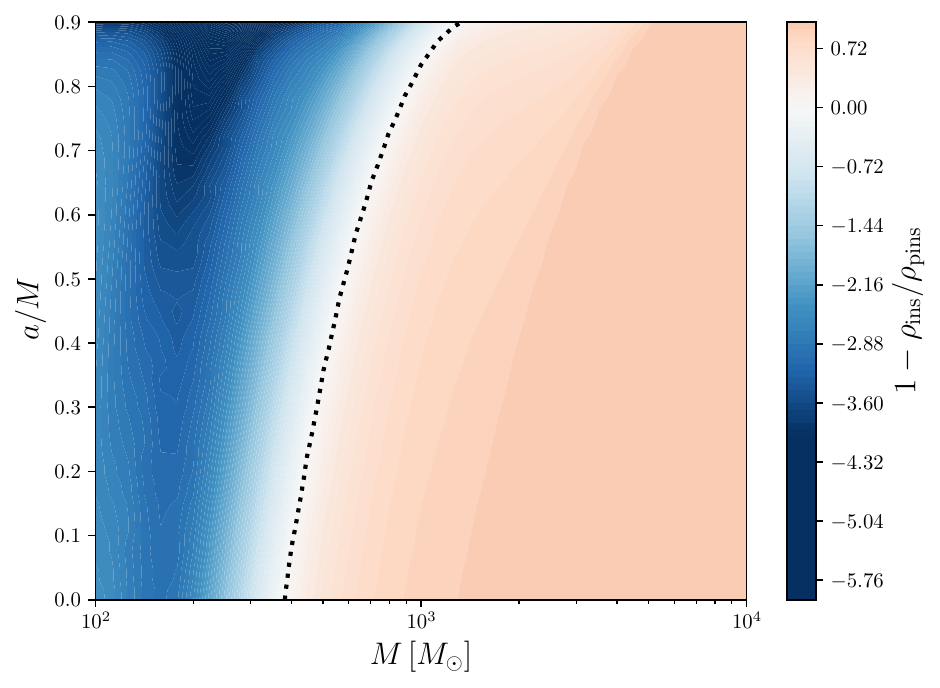}
    \caption{
    Contour map of the relative difference between the inspiral and post-inspiral SNRs, 
    $ 1 - \rho_\mathrm{ins} / \rho_\mathrm{pins}$, 
    for intermediate-mass-ratio binary BH mergers observed with ET. 
    The dotted line indicates where the two SNRs are equal, $\rho_\mathrm{ins} = \rho_\mathrm{pins}$.}
    \label{fig:SNR_fraction}
\end{figure}

Figure~\ref{fig:strain_plot} shows the averaged frequency-domain strain 
$2\sqrt{f}|\tilde{h}_\mathrm{ave}(f)|~\mathrm{[Hz^{-1/2}]}$ of a representative intermediate-mass-ratio binary BH merger (blue) together with the square root of the noise PSD $\sqrt{S_\mathrm{n}(f)}~\mathrm{[Hz^{-1/2}]}$ (black dashed). 
The vertical line marks the ISCO frequency, which separates the inspiral and post-inspiral regimes. 
Corresponding SNRs in the two regimes are $\rho_\mathrm{ins}=30$ and $\rho_\mathrm{pins}=111$ respectively.

Fixing mass ratio and luminosity distance, a key parameter that determines the relative strengths of $\rho_\mathrm{ins}$ and $\rho_\mathrm{pins}$ is the ISCO frequency $f_\mathrm{ISCO}$.
Since $f_\mathrm{ISCO} \propto 1/M$ for a fixed $a/M$, a larger primary mass $M$ shifts the entire signal to lower frequencies, reducing the overlap of the inspiral waveform with the most sensitive band of ET and thereby decreasing $\rho_\mathrm{ins}$. 
On the other hand, for larger primary spin $a$ with a fixed $M$, $f_\mathrm{ISCO}$ increases, extending the inspiral part of the signal into the high-sensitivity region of the detector, which enhances $\rho_\mathrm{ins}$. 
The competition between these two effects, mass-driven reduction and spin-driven enhancement, essentially determines the ratio $\rho_\mathrm{ins}/\rho_\mathrm{pins}$ across the parameter space.

Figure~\ref{fig:SNR_fraction} shows the contour map of the fractional difference $1-\rho_\mathrm{ins}/\rho_\mathrm{pins}$. 
The dotted line marks where $\rho_\mathrm{ins} = \rho_\mathrm{pins}$, determined by the ISCO frequency $f_\mathrm{ISCO}$, which depends on $M$ and $a$.  
The optimal primary mass $M$, for which the inspiral and post-inspiral SNRs become comparable, lies between $4\times10^2\,M_\odot$ and $10^3\,M_\odot$, and shifts toward higher values as $a$ increases.  
To the left of the contour (smaller $M$), the inspiral phase dominates the total SNR, whereas to the right (larger $M$), the post-inspiral contribution becomes dominant.
In Appendix~\ref{app:ins_pins_snrs}, we show the contour maps of $\rho_\mathrm{ins}$ and $\rho_\mathrm{pins}$ over the parameter space in Fig.~\ref{fig:ins_post_ins_SNR}.

The region where both the inspiral and post-inspiral SNRs are comparable is an ideal target for the inspiral-merger-ringdown consistency test, which compares the remnant mass and spin estimated independently from the inspiral and post-inspiral signals assuming GR~\cite{LIGOScientific:2016vlm, Ghosh:2016qgn, Ghosh:2017gfp, LIGOScientific:2019fpa, LIGOScientific:2020tif, LIGOScientific:2021sio, Madekar:2024zdj, Shaikh:2024wyn}.
If the individual SNRs are sufficiently high, GW signals from such systems will not only enable stringent tests of the internal consistency of GR, as in previous analyses, but also allow separate tests of gravity.  
For example, the inspiral regime can be analyzed within the parametrized post-Einsteinian framework, in which beyond-GR effects are introduced as perturbative corrections to the PN waveform (e.g.~\cite{Yunes:2009ke, Tahura:2018zuq}).
Similarly, the ringdown regime can be inferred using parametrized quasi-normal-mode models that incorporate beyond-GR modifications to the quasi-normal-mode spectrum (e.g.~\cite{Hirano:2024fgp, Cano:2024jkd, Chung:2024ira}). 
By comparing the constraints obtained independently from these two regimes, we can perform more robust tests of deviations from GR and/or the Kerr geometry.

In the region where the post-inspiral SNR dominates, waveform models that terminate at the ISCO are no longer sufficient for extracting the full information contained in the signal.
By explicitly modeling the post-ISCO dynamics, our framework enables these parts of the parameter space to be incorporated into data analysis rather than discarded.
This substantially broadens the range of systems for which accurate parameter estimation can be performed, particularly for higher primary masses and larger spins, where the post-inspiral contribution carries the majority of the detectable SNR.
In the next subsection, we demonstrate that incorporating the post-inspiral regime not only enlarges the usable parameter space but also enables more precise measurements of the BH charge.

\subsection{Distinguishability of the BH charge}
\label{sec:distinguishability_IMR}
In this subsection, we quantify how the inclusion of the post-inspiral portion of the waveform improves the constraints on the BH charge.  
To do so, we evaluate the mismatch between the waveforms for the charged and neutral (Kerr) primary BH over the inspiral, post-inspiral, and full inspiral–transition–plunge frequency bands.

\subsubsection{\textbf{Waveform mismatch}}
\label{sec:waveform_mismatch}
To quantify the impact of the BH charge on the waveform and to assess the benefit of modeling the inspiral, transition, and plunge, we compute the mismatch between the charged waveform and the Kerr case.
For a given frequency interval $(f_1, f_2)$, we define the mismatch as
\begin{equation}
\begin{split}
    &\mathcal{MM}(Q/M; f_1, f_2)\\
    &:= 1 - 
    \max_{t_0,\phi_0}
    \frac{\langle h^{\mathrm{Kerr}}_{22} \, \big| \, h_{22}(Q/M) \mathrm{e}^{\mathrm{i}\phi_0 - 2\pi \mathrm{i} f t_0} \rangle}
    {\sqrt{\langle h^{\mathrm{Kerr}}_{22} | h^{\mathrm{Kerr}}_{22} \rangle\;\langle h_{22}(Q/M) | h_{22}(Q/M) \rangle}}\:,
\end{split}
\end{equation}
where each inner product is evaluated over a frequency range $f_1\leq f \leq f_2$, $h_{22}(Q/M)$ is a waveform with non-zero charge $Q$ for the primary BH, and $h^\mathrm{Kerr}_{22}$ is the corresponding waveform for the Kerr primary with zero charge.
The maximization over $(t_0,\phi_0)$ accounts for arbitrary relative time and phase shifts between the two waveforms.
Using this definition, we introduce the inspiral (ins), post-inspiral (pins), and inspiral–transition–plunge (ITP) mismatches as
\begin{align}
    \mathcal{MM}_{\mathrm{ins}}(Q/M)
    &:= \mathcal{MM}\left(Q/M;\, f_\mathrm{min},\, f_\mathrm{ISCO}\right), \\
    \mathcal{MM}_{\mathrm{pins}}(Q/M)
    &:= \mathcal{MM}\left(Q/M;\, f_\mathrm{ISCO},\, f_\mathrm{max}\right), \\
    \mathcal{MM}_{\mathrm{ITP}}(Q/M)
    &:= \mathcal{MM}\left(Q/M;\, f_\mathrm{min},\, f_\mathrm{max}\right).
\end{align}

\subsubsection{\textbf{Distinguishability criteria}}
\label{sec:distinguishability_criteria}
There is a useful criterion connecting the SNR to the mismatch $\mathcal{MM}$ that can be distinguished in parameter estimation
(e.g.,~\cite{Chatziioannou:2017tdw}):
\begin{equation}
    \mathcal{MM}< \frac{N}{2 \rho^2}\:,
    \label{eq:mismatch_snr}
\end{equation}
where $N$ denotes the number of parameters in the waveform model.
In our case, the waveform model contains $D=11$ parameters: four intrinsic parameters describing the binary—the primary mass, spin, charge, and mass ratio—together with seven extrinsic parameters (luminosity distance, sky location, inclination angle, polarization angle, and the reference time and phase).

Using Eq.~\eqref{eq:mismatch_snr}, we define the corresponding mismatch 
threshold for a given SNR $\rho$ as
\begin{equation}
    \mathcal{MM}^{\mathrm{(thr)}} := \frac{N}{2\rho^2}\:.
\end{equation}
For quantities evaluated over restricted frequency bands, such as the inspiral, post-inspiral, or full inspiral–transition–plunge regimes, we denote the respective thresholds by $\mathcal{MM}_{\mathrm{ins}}^{\mathrm{(thr)}}, \mathcal{MM}_{\mathrm{pins}}^{\mathrm{(thr)}}, \mathcal{MM}_{\mathrm{ITP}}^{\mathrm{(thr)}}$, where each mismatch is computed using the SNR accumulated in the corresponding frequency range.
We then translate these threshold mismatches into a critical value of the charge by defining $|Q|^{\mathrm{(thr)}}/M$ as the 
value of $|Q|/M$ that satisfies
\begin{equation}
    \label{eq:mismatch_snr_thr}
    \mathcal{MM}\left(\frac{|Q|}{M}\right)
    = \mathcal{MM}^{\mathrm{(thr)}}\:,
\end{equation}
and, analogously, $|Q|^{\mathrm{(thr)}}_{\mathrm{ins}}/M$,
$|Q|^{\mathrm{(thr)}}_{\mathrm{pins}}/M$, and
$|Q|^{\mathrm{(thr)}}_{\mathrm{ITP}}/M$ for the inspiral, post-inspiral, 
and ITP regimes, respectively.
These critical charges indicate the minimal $|Q|/M$ for which the deviation from the Kerr waveform becomes distinguishable in each case.

\begin{figure}
    \centering
    \includegraphics[width=\linewidth]{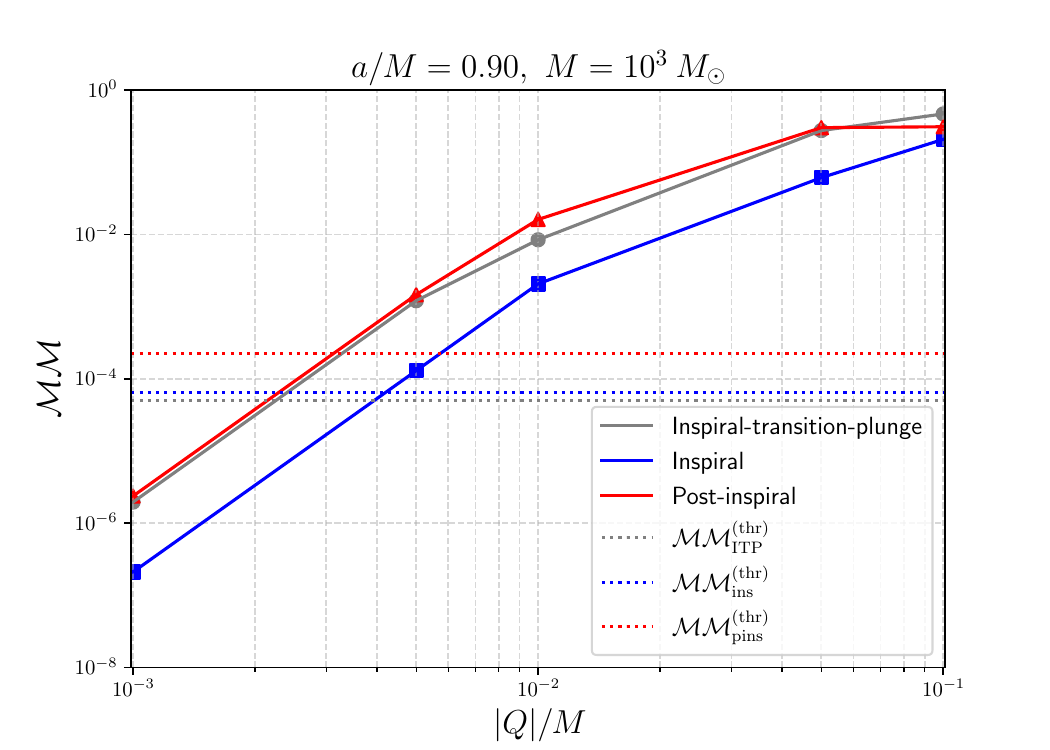}
    \caption{Mismatch $\mathcal{MM}$ between the waveform with a charged primary and the Kerr case for the BH spin $a/M = 0.90$, the mass ratio $\eta=10^{-2}$, the detector-frame mass of  the primary BH $M=10^{3}\,M_{\odot}$, the distance  $D=5$ Gpc, and using ET.
    The solid curves show the mismatches computed using the inspiral $\mathcal{MM}_\mathrm{ins}$ (blue), the post-inspiral $\mathcal{MM}_\mathrm{pins}$ (red), and the full inspiral–transition–plunge waveforms $\mathcal{MM}_\mathrm{ITP}$ (gray). 
    The horizontal dashed lines indicate the distinguishability thresholds $\mathcal{MM}^\mathrm{(thr)}_{\mathrm{ITP}}$, $\mathcal{MM}^\mathrm{(thr)}_{\mathrm{ins}}$, and 
    $\mathcal{MM}^\mathrm{(thr)}_{\mathrm{pins}}$, derived from Eq.~\eqref{eq:mismatch_snr_thr} for the corresponding SNRs. 
    Notice that the mismatch for the post-inspiral is always larger than that for the inspiral for the configuration considered here, indicating that the post-inspiral signal is more sensitive to $Q/M$ than the inspiral one.}
    \label{fig:mismatch_a0.9_logM3}
\end{figure}

\subsubsection{\textbf{Results}}
Figure~\ref{fig:mismatch_a0.9_logM3} shows how the mismatch 
$\mathcal{MM}$ between the waveform with a charged primary and the Kerr case varies as a function of $|Q|/M$ for $a/M=0.90$, $\eta=10^{-2}$, $M = 10^{3} M_{\odot}$, and $D = 5~\mathrm{Gpc}$ using ET.
The inspiral, post-inspiral, and full inspiral–transition–plunge 
mismatches---$\mathcal{MM}_{\mathrm{ins}}$, $\mathcal{MM}_{\mathrm{pins}}$, 
and $\mathcal{MM}_{\mathrm{ITP}}$---are shown by the blue, red, and gray solid curves, respectively.
For this configuration, $\mathcal{MM}_{\mathrm{pins}}$ exceeds 
$\mathcal{MM}_{\mathrm{ins}}$ across the range of $|Q|/M$, implying that the post-inspiral regime is more sensitive to the presence of the charge for a fixed SNR.
The horizontal dashed lines indicate the corresponding mismatch thresholds 
$\mathcal{MM}^{\mathrm{(thr)}}_{\mathrm{ins}}$, 
$\mathcal{MM}^{\mathrm{(thr)}}_{\mathrm{pins}}$, and 
$\mathcal{MM}^{\mathrm{(thr)}}_{\mathrm{ITP}}$, computed from Eq.~\eqref{eq:mismatch_snr_thr}.
By identifying where each solid curve intersects its respective threshold, we obtain the critical charge values
\begin{align}
    |Q|^{\mathrm{(thr)}}_{\mathrm{ins}}/M &= 4.2\times10^{-3}\:,\\
    |Q|^{\mathrm{(thr)}}_{\mathrm{pins}}/M &= 3.1\times10^{-3}\:,\\    |Q|^{\mathrm{(thr)}}_{\mathrm{ITP}}/M &= 2.3\times10^{-3}.
\end{align}
These quantify the minimum charge-to-mass ratio for which the deviation from the Kerr waveform becomes distinguishable in parameter estimation.
Overall, the post-inspiral regime provides stronger constraints on $Q/M$ than the inspiral regime, while modeling the entire inspiral-transition-plunge waveform yields sensitivity to even smaller values of $|Q|/M$.

\begin{figure*}
    \centering
    \includegraphics[width=\linewidth]{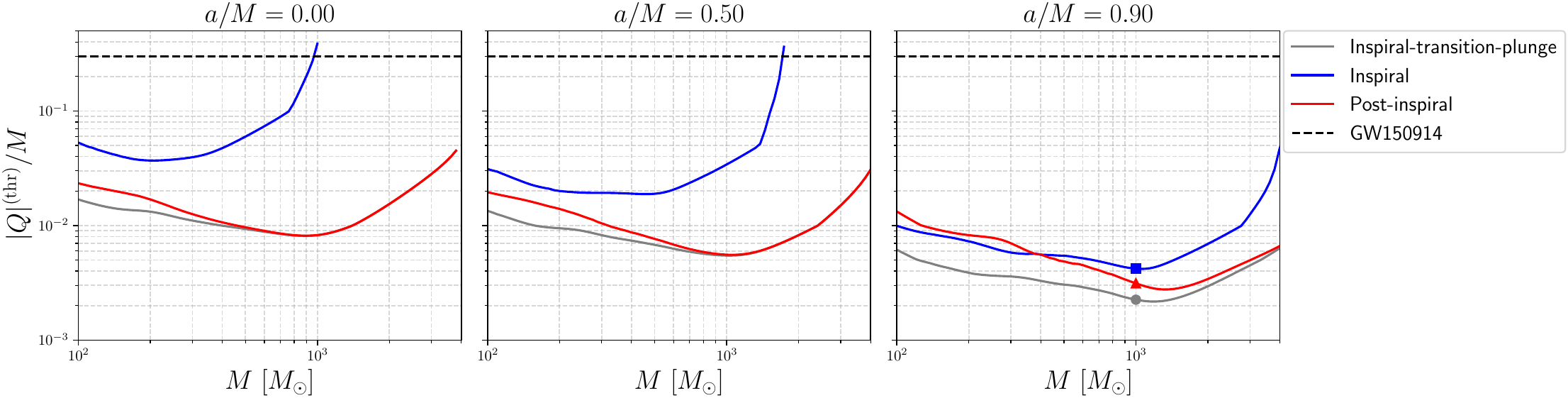}
    \caption{
    Threshold charge-to-mass ratios $|Q|^{\mathrm{(thr)}}/M$ as functions of the detector-frame primary mass $M$ for three primary's spin values, $a/M = 0$, $0.5$, and $0.9$. 
    The curves show the critical charge required for distinguishability based on the mismatch criterion in Eq.~\eqref{eq:mismatch_snr_thr}, computed separately from the inspiral (blue), post-inspiral (red), and full inspiral–transition–plunge (gray) waveform segments.  
    The horizontal black dashed line indicates the threshold charge for GW150914 $|Q|/M = 0.3$~\cite{Bozzola:2020mjx, Carullo:2021oxn}, shown for reference.
   The points highlighted at $M=10^3M_\odot$ in the right panel correspond to $|Q|^\mathrm{(thr)}/M$ shown in Fig.~\ref{fig:mismatch_a0.9_logM3}.}
    \label{fig:Q_thr}
\end{figure*}

Let us now evaluate the critical charge values for various  primary BH mass $M$ for three different spins: $a/M = 0$, $0.5$, and $0.9$. 
We restrict our analysis to the mass range $100 M_{\odot} \leq M \leq 4000 M_{\odot}$.  
For $M \gtrsim 4000 M_{\odot}$, the SNR rapidly decreases to a level at which the mismatch indicator $\mathcal{MM}^{\mathrm{(thr)}}$ is no longer reliable (see Fig.~\ref{fig:ins_post_ins_SNR}).  
Therefore, we do not include such high masses in our plots.

Figure~\ref{fig:Q_thr} shows the threshold charge-to-mass ratios 
$|Q|^{\mathrm{(thr)}}/M$ obtained from the inspiral, post-inspiral, and full inspiral–transition–plunge waveform segments as a function of $M$.  
Lower values of $|Q|^{\mathrm{(thr)}}/M$ correspond to stronger sensitivity to the presence of charge.
Across all spins, the post-inspiral regime (red curves) typically yields smaller $|Q|^{\mathrm{(thr)}}/M$ than the inspiral regime (blue curves), indicating that the post-inspiral signal provides better sensitivity to the BH charge for a fixed $M$.  
The inspiral–transition–plunge threshold (gray), which incorporates the full waveform, offers the most stringent constraints over the entire mass range.  
For rapidly spinning primaries ($a/M = 0.9$), the detectable charge improves further when using either the post-inspiral segment or the full waveform, 
reaching values below $|Q|/M \sim 3 \times 10^{-3}$ around $M \approx 10^{3} M_{\odot}$.
The horizontal dashed line indicates the threshold charge inferred from GW150914~\cite{Bozzola:2020mjx, Carullo:2021oxn}, shown for comparison.  
A detection of an intermediate–mass–ratio binary merger with ET is likely to improve that bound on the BH charge by 1 to 2 orders of magnitude compared to the current bound from GW150914, especially when the post-inspiral signal is included in the analysis.

\subsection{Accumulated phase difference in extreme mass-ratio inspirals}
\label{sec:EMRI}

Up to this point, our analysis has focused on intermediate–mass–ratio binary mergers, where both the inspiral and post-inspiral regimes contribute significantly to the observed signal.  
We now turn to the extreme–mass–ratio case, in which the inspiral overwhelmingly dominates the waveform and even small modifications to the dynamics can accumulate into large dephasings over the long inspiral duration.

As a key ingredient for accurately modeling extreme–mass–ratio inspirals (EMRIs), Sec.~\ref{sec:energy_flux} presented the GW energy flux from a circular orbit.  
This flux was computed by accounting for both the outgoing radiation and the ingoing flux absorbed by the horizon, using the Teukolsky–Starobinsky identity for the Kerr--Newman spacetime (Eq.~\eqref{eq:TS-id-KN}).
Incorporating this full flux is essential for capturing the inspiral evolution with sufficient precision to assess the impact of a nonzero charge on the accumulated phase.
With this flux calculation in hand, we now evaluate the resulting dephasing in EMRIs and quantify how sensitively such systems can probe deviations from the Kerr geometry.

We consider an EMRI with the detector-frame primary mass $(1+z)M=10^6~M_\odot$ and $\eta=10^{-5}$, starting from $r=10M$. 
Since the GW phase of the $(\ell, m)$ mode is related to the orbital phase by $\phi_{\ell m} = m\,\phi_\mathrm{orb}$ in the inspiral stage, the minimum frequency of the dominant mode is $2.04~\mathrm{mHz}$ for $a/M=0$, $2.01~\mathrm{mHz}$ for $a/M=0.5$, and $1.99~\mathrm{mHz}$ for $a/M=0.9$, all of which lie well within the sensitivity bands of the detectors.

\subsubsection{\textbf{Distinguishability criteria}}

To assess the distinguishability of the BH charge, we adopt the accumulated 
phase difference from the Kerr case rather than the waveform mismatch.  
This choice is motivated both by computational practicality—the phase difference is far easier to evaluate than the full mismatch—and by the fact that accumulated dephasing is a standard and widely used diagnostic of EMRI waveform accuracy (e.g.~\cite{Burke:2023lno, Khalvati:2025znb}).

For $a/M = 0$ and $0.5$, the inspiral from $r = 10M$ to the ISCO occurs within four years.
In these cases, we evaluate the accumulated GW phase at the ISCO as
\begin{equation}
    \Phi_{22}(Q/M)
    := 2\bigl[\phi_{\mathrm{orb}}(r_{\mathrm{ISCO}}) - \phi_{\mathrm{orb}}(10M)\bigr].
\end{equation}
For $a/M = 0.9$, however, the inspiral time from $r = 10M$ to the ISCO exceeds four years.
In this case, we instead evaluate the accumulated phase over the final four years before ISCO,
\begin{equation}
    \Phi_{22}(Q/M)
    := 2\bigl[\phi_{\mathrm{orb}}(t_{\mathrm{ISCO}}) - \phi_{\mathrm{orb}}(t_{\mathrm{ISCO}} - 4~\mathrm{yr})\bigr],
\end{equation}
where $t_{\mathrm{ISCO}}$ denotes the coordinate time at which the orbit reaches the ISCO.
We adopt the conservative criterion that two waveforms are distinguishable if the accumulated phase difference $\Delta \Phi_{22}(Q/M)$ exceeds one radian,
$\Delta\Phi_{22} \ge 1~\mathrm{radian}$~\cite{Lindblom:2008cm, Bonga:2019ycj, Burke:2023lno, Khalvati:2025znb}.

\subsubsection{\textbf{Results}}
\begin{figure}
    \centering
    \includegraphics[width=1.\linewidth]{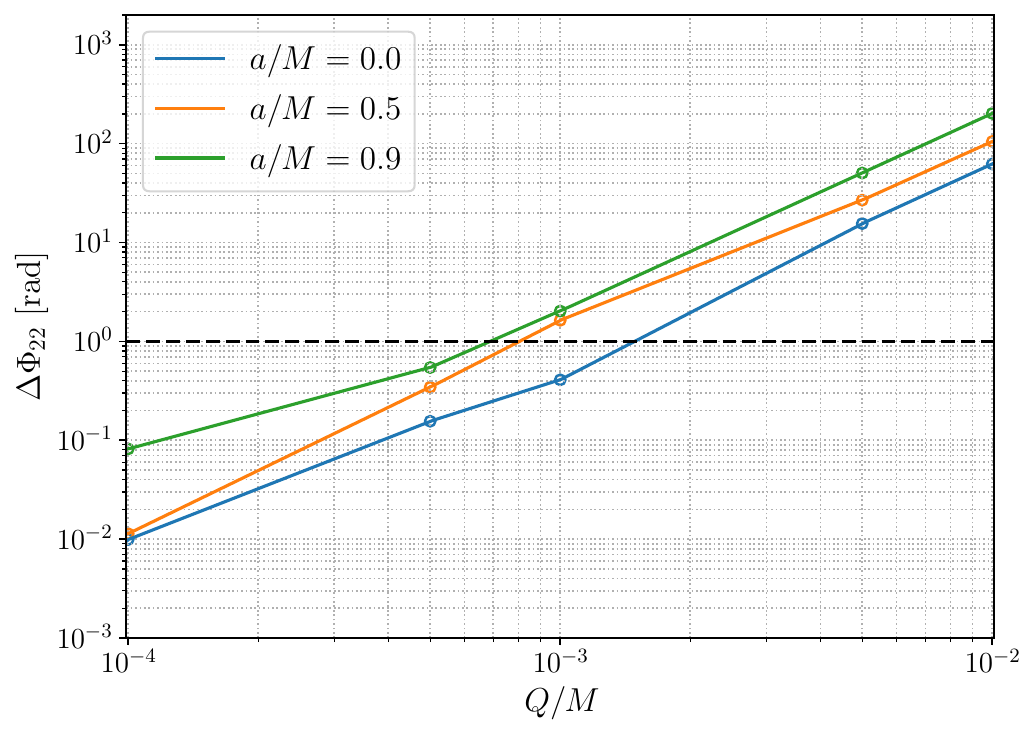}
    \caption{
    The accumulated phase differences $\Delta\Phi_{22}$ as a function of $Q/M$ for the mass ratio $\eta = 10^{-5}$ with the BH spin $a/M = 0$, $0.5$, and $0.9$, shown in blue, orange, and green, respectively. 
    The circles indicate the computed data points, and the solid lines are obtained by linear interpolation in $Q/M$ on a logarithmic scale. 
    By adopting the criterion that $\Delta\Phi_{22}(Q/M) \geq 1~\mathrm{rad}$ is distinguishable with space-based GW interferometers, 
    the corresponding thresholds are $|Q|/M \geq 1.5\times10^{-3}$, $8.0\times10^{-4}$, and $6.9\times10^{-4}$ for $a/M = 0$, $0.5$, and $0.9$, respectively.}
    \label{fig:phase_diff}
\end{figure}

Figure~\ref{fig:phase_diff} shows the accumulated phase difference $\Delta\Phi_{22}(Q/M)$ for $a/M = 0$, $0.5$, and $0.9$. 
Since the phase shift induced by the charge is expected to scale as $|Q|/M$ in the small-charge regime ($|Q|/M \ll 1$), we perform a linear interpolation in $|Q|/M$ on a logarithmic scale to estimate the detection threshold. 
The results indicate that deviations become distinguishable for $|Q|/M \geq 1.5\times10^{-3}$, $8.0\times10^{-4}$, and $6.9\times10^{-4}$ 
for $a/M = 0$, $0.5$, and $0.9$, respectively.

These thresholds are consistent in order of magnitude with the results of a recent Fisher-matrix analysis~\cite{Zi:2022hcc}, which implemented a charged secondary in an analytic–kludge EMRI waveform around a Kerr--Newman BH and evaluated the measurability of the BH charge using the expected sensitivity of the TianQin detector. 

\section{Discussion}
\label{sec:discussion}
In this section, we summarize and comment on the major assumptions and simplifications adopted in our calculations.

\begin{itemize}
    \item \textbf{Neglecting the electromagnetic perturbation (Dudley–Finley approximation)}\\
    Given that the primary BH charge is small ($|Q|/M \ll 1$), and the secondary is neutral, the electromagnetic perturbation induced in the Kerr--Newman background is expected to be extremely weak. 
    We therefore set the electromagnetic perturbation to zero and focus solely on the gravitational sector. 
    This treatment corresponds to the Dudley–Finley approximation, in which one of the two coupled sectors (gravitational or electromagnetic) is held fixed to simplify the system.

    In the computation of quasi-normal mode frequencies, the Dudley--Finley approximation has been shown to yield accurate results~\cite{Berti:2005eb, Saha:2025nsg} for a sufficiently small BH charge.
    A recent analysis~\cite{Saha:2025nsg}, which directly compared the Dudley--Finely approximation with results obtained from the fully coupled gravito–electromagnetic perturbation equations~\cite{Dias:2015wqa, Dias:2021yju, Carullo:2021oxn, Dias:2022oqm}, demonstrated that the real and imaginary parts of the QNM frequencies agree within $\lesssim 10~\%$ and $\lesssim 1~\%$, respectively, across parameter space including the near-extremal regime.
    Moreover, for small charge-to-mass ratios ($|Q|/M \ll 1$), the discrepancies are further reduced, typically remaining $\lesssim 0.1~\%$.

    Motivated by this demonstrated accuracy, we adopt the same approximation in the present analysis, assuming that possible $\mathcal{O}(Q^2)$ effects such as dipolar electromagnetic radiation remain subdominant in the presence of the source.  
    Extending the analysis to include the fully coupled system would allow us to quantify these effects more accurately.

    \vspace{0.3em}
    \item \textbf{Considering a neutral, non-spinning secondary orbiting a Kerr--Newman BH}\\

    In this work, we restrict our attention to a system consisting of a charged, rotating primary and a neutral, non-spinning point particle. 
    Because the secondary is neutral, there is no direct electromagnetic radiation sourced by the particle itself, and neglecting this channel is therefore a reasonable approximation in the present analysis.
    
    Nevertheless, in more general setups where the secondary carries charge and/or spin, additional effects such as direct electromagnetic emission, spin--orbit coupling, enhanced gravito–electromagnetic coupling, and electric force between the two BHs are expected to arise. 
    For the inspiral phase, such charge-induced corrections have already been incorporated within analytic–kludge EMRI models around Kerr--Newman BHs, where both the GW and dipolar electromagnetic radiation channels are included in the orbital evolution~\cite{Zi:2022hcc}. 
    Extending the waveform modeling to the full inspiral–transition–plunge regime while consistently accounting for these charge- and spin-induced effects will be essential for producing more realistic templates for charged binaries and for assessing their observational signatures.
    A natural next step is to generalize our framework to generic (inclined and eccentric) orbits.  
    Such an extension would enable a more complete description of realistic asymmetric-mass systems and provide an even more powerful setting for testing deviations from the Kerr paradigm.

\end{itemize}

\section{Conclusion}
\label{sec:conclusion}
In this work, we calculated, for the first time, GWs radiated from a highly asymmetric-mass binary during its entire evolution of the inspiral, transition, and plunge stages of a merger involving a BH beyond Kerr, taking the Kerr--Newman BH as a concrete example
We followed the Teukolsky formalism and assumed the secondary to be neutral and non-spinning as a first step calculation.
We first constructed the secondary's continuous worldline, covering these three regimes, following the procedure developed in AH19 for highly asymmetric-mass binaries with a Kerr BH.
Inserting such constructed worldline into the perturbation equation, we calculated the GWs for various primary charges and spins. 

We also studied observational prospects. 
For intermediate–mass–ratio binaries observed with ET, we found that the post-inspiral portion of the signal can dominate the detectable power once the primary mass exceeds $M \sim 400$--$1000\,M_\odot$ while fixing the mass ratio to $10^{-2}$ and the luminosity distance to $5~\mathrm{Gpc}$.
The crossover between inspiral-dominated and post-inspiral–dominated regimes shifts to larger primary masses for higher spins.
In the vicinity of the crossover between inspiral-dominated and post-inspiral–dominated regimes, the signal contains comparable information from both phases, making these systems ideal targets for inspiral–merger–ringdown consistency–type tests.
In the regime where the post-inspiral part dominates, waveform models that focus solely on the inspiral would omit a significant fraction of the signal, whereas our inspiral–transition–plunge framework enables full recovery of the post-inspiral information.  
A mismatch-based analysis further showed that the post-inspiral phase provides tighter constraints on the charge-to-mass ratio than the inspiral alone, and that combining all phases yields the strongest sensitivity, reaching $|Q|/M \sim \mathrm{a~few}\times 10^{-3}$ for rapidly spinning primaries near $M = 10^{3}\,M_\odot$. 
In general, we expect observations of intermediate-mass-ratio mergers with ET to improve the bound on the BH charge by one to two orders of magnitude from the current bound with GW150914, especially when the post-inspiral signal is included in the analysis.

For extreme–mass–ratio inspirals (EMRIs) observed by space-based detectors, such as LISA, the longer duration of the inspiral allows even small deviations from the Kerr geometry to accumulate into measurable phase differences.  
Using the Kerr--Newman fluxes derived here, we found that charges as small as 
$|Q|/M \sim 10^{-3}$--$10^{-4}$ would lead to distinguishable dephasings over a four-year observation, depending on the primary spin, highlighting the exceptional sensitivity of EMRIs to small departures from Kerr.

Several avenues exist for future works.
For example, one could include the secondary's charge and account for the electromagnetic radiation as well as the electric force, which will modify the binary evolution and GWs from the current work, where we ignored the secondary's charge.
In particular, it would be important to relax the Dudley-Finley approximation adopted here and treat both electromagnetic and gravitational perturbations consistently, including the coupling between these two perturbations. Another important extension to include secondary's spin.
Furthermore, it would be interesting to consider more generic orbits, such as non-equatorial or those with non-vanishing eccentricity.
Finally, it would be interesting to construct inspiral-transition-plunge waveforms in theories beyond General Relativity (see~\cite{Roy:2025kra} for a recent work that constructed post-merger waveforms for highly asymmetric binaries in higher curvature theories of gravity using a self-force framework).

\begin{acknowledgments}
We gratefully acknowledge Scott A. Hughes for helpful discussions on the collocation time between the \ins\ and \tra. 
We thank Hidetoshi Omiya and Takahiro S. Yamamoto for fruitful discussions and comments.
This work makes use of the Black Hole Perturbation Toolkit. 
D.~W. is supported by JSPS KAKENHI grant No. 23KJ06945.
K.~Y. acknowledges support from the NSF grant PHY-2309066 and the NSF-CAREER Award PHYS-2339969.
\end{acknowledgments}

\appendix
\begin{widetext}
\section{Source term of the Dudley--Finley equation}
\label{app:source_term}

In this appendix, we present lengthy expressions for the source term of the Dudley--Finley equation. 
The form is identical to that of the Teukolsky equation~\cite{Sasaki:2003xr}, and the dependence of $Q$ appears only through the function $\Delta$ for the case of the vanishing secondary charge considered here.

First, $B'_2$ and $B'^*_2$ in Eq.~\eqref{eq:T_lmw} are given by
\begin{align}
B'_2 &= -\frac{1}{2} \rho^8 \Bar{\rho} \hat{L}_{-1} \left[ \rho^{-4} \hat{L}_0 \left( \rho^{-2} \Bar{\rho}^{-1} T_{nn} \right) \right] -\frac{1}{2\sqrt{2}} \rho^8 \Bar{\rho} \Delta^2 \hat{L}_{-1} \left[ \rho^{-4} \Bar{\rho}^2 \hat{J}_+ \left( \rho^{-2} \Bar{\rho}^{-2} \Delta^{-1} T_{mn} \right) \right]\:,\\
B'^*_2 &= -\frac{1}{4} \rho^8 \Delta^2 \hat{J}_+ \left[ \rho^{-4} \Bar{\rho}^2 \hat{J}_+ \left( \rho^{-2} \Bar{\rho} T_{mm} \right) \right] -\frac{1}{2\sqrt{2}} \rho^8 \Delta^2 \hat{J}_+ \left[ \rho^{-4} \Bar{\rho}^2 \Delta^{-1} \hat{L}_{-1} \left( \rho^{-2} \Bar{\rho}^{-2} T_{mn} \right) \right]\:,
\end{align}
where 
\begin{align}
    \hat{L}_s  &= \frac{\partial}{\partial \theta} - \frac{\mathrm{i}}{\sin\theta} \frac{\partial}{\partial \phi} - \mathrm{i}\sin\theta\frac{\partial}{\partial t} + s\cot\theta\:,\\
    \hat{J}_+ &= \frac{\partial}{\partial r} - \left( \frac{r^2+a^2}{\Delta}\frac{\partial}{\partial t} + \frac{a}{\Delta}\frac{\partial}{\partial \phi} \right)\:.
\end{align}
$T_{nn}, T_{\overline{m}n},$ and $T_{\overline{mm}}$ are the tetrad components of the energy-momentum tensor, which is given by
\begin{equation}
\begin{split}
    T^{\alpha\beta} = &\frac{\mu}{\Sigma\sin\theta \frac{\diff t}{\diff \tau}} \frac{\diff z^\alpha}{\diff \tau} \frac{\diff z^\beta}{\diff \tau} \delta \left( r-r(t) \right) \delta \left( \theta-\theta(t) \right) \delta \left( \phi-\phi(t) \right)\:,
\end{split}
\end{equation}
where $z^\alpha=(t, r(t), \theta(t), \phi(t))$ is the secondary's trajectory. 
For a geodesic motion, explicit forms of $T_{nn}, T_{\overline{m}n},$ and $T_{\overline{mm}}$ are given by 
\begin{equation}
    T_{AB} = \mu \frac{C_{AB}}{\sin\theta} \delta \left( r-r(t) \right) \delta \left( \theta-\theta(t) \right)  \delta \left( \phi-\phi(t) \right)\:,
\end{equation}
where the indices $A$ and $B$ take either $n$ or $\bar m$ while the coefficients $C_{AB}$ are given by
\begin{align}
    C_{nn} &= \frac{1}{4\Sigma^3} \left( \frac{\diff t}{\diff \tau} \right)^{-1} \left( P + \Sigma \frac{\diff r}{\diff \tau} \right)^2\:,\\
    C_{\overline{m}n} &= -\frac{\rho}{2\sqrt{2} \Sigma^2} \left( \frac{\diff t}{\diff \tau} \right)^{-1} \left( P + \Sigma \frac{\diff r}{\diff \tau} \right)\left[\mathrm{i}\sin\theta \left(a\hat{E}-\frac{\hat{L}}{\sin\theta}\right)+\Sigma\frac{\diff \theta}{\diff \tau}\right] \:,\\
    C_{\overline{mm}} &= -\frac{\rho^2}{2\Sigma} \left( \frac{\diff t}{\diff \tau} \right)^{-1} \left[ \mathrm{i}\sin\theta\left(a\hat{E}-\frac{\hat{L}}{\sin\theta}\right) + \Sigma\frac{\diff \theta}{\diff \tau}\right]^2\:.
\end{align}

Next, let us present expressions for $T_{\ell m\omega}$.
Using an identity for arbitrary functions $a(\theta)$ and $b(\theta)$
\begin{equation}
    \int^{\pi}_0 a(\theta) L_s[b(\theta)] \sin\theta \diff \theta = -\int^\pi_0 b(\theta) L^\dagger_{1-s}[a(\theta)] \sin\theta \diff \theta\:,
\end{equation}
we can rewrite Eq.~\eqref{eq:T_lmw} as
\begin{equation}
\label{eq:T_lmw_2}
    \begin{split}
        T_{\ell m\omega} =\:\frac{4\mu}{\sqrt{2\pi}} &\int^\infty_{-\infty} \diff t \int \diff \theta \:\mathrm{e}^{\mathrm{i}\omega t -\mathrm{i}m\phi(t)}\\
        &\:\times\Bigg\{ -\frac{1}{2}L^\dagger_1\left[ \rho^{-4} L^\dagger_2 (\rho^3 S) \right] C_{nn}\rho^{-2}\Bar{\rho}^{-1} \delta(r-r(t))\delta(\theta-\theta(t)) \\
        &\:\:\:\:\:\:\:\:\:\:+\frac{\Delta^2 \bar{\rho}}{\sqrt{2}\rho^2} L^\dagger_2 (\rho\Bar{\rho}S) J_+\left[ \frac{C_{\overline{m}n}}{\rho^2\Bar{\rho}^2\Delta}\delta(r-r(t))\delta(\theta-\theta(t)) \right] \\
        &\:\:\:\:\:\:\:\:\:\:+\frac{1}{2\sqrt{2}}L^\dagger_{2}\left[ \rho^3 S\frac{\partial }{\partial r}(\Bar{\rho}^2 \rho^{-4})\right] \frac{C_{\overline{m}n}\Delta}{\rho^{2}\Bar{\rho}^{2}} \delta(r-r(t))\delta(\theta-\theta(t))\\
        &\:\:\:\:\:\:\:\:\:\:-\frac{1}{4}\rho^3 \Delta^2 S J_+\left[ \rho^{-4}J_+(\Bar{\rho}\rho^{-2} C_{\overline{mm}}\delta(r-r(t))\delta(\theta-\theta(t)) ) \right]
        \Bigg\}\:,
    \end{split}
\end{equation}
where
\begin{align}
    &L^\dagger_s = \frac{\partial}{\partial \theta} - \frac{m}{\sin\theta} + a\omega \sin\theta + s\cot\theta\:,\\
    & J_+ = \frac{\partial}{\partial r} + \mathrm{i}\frac{K}{\Delta}\:.    
\end{align}
For a source with compact support, we can perform a partial integral and derive
\begin{equation}
\begin{split}
    T_{\ell m\omega} = \mu \int^{\infty}_\infty \diff t~\mathrm{e}^{\mathrm{i}\omega t-\mathrm{i}m\phi(t)} \Delta^2 \Big\{ &(A_{nn0} + A_{\bar{m}n0} + A_{\bar{m}\bar{m}0})\delta(r-r(t))\\ &+ \left[(A_{\bar{m}n1}+A_{\bar{m}\bar{m}1})\delta(r-r(t))\right]_{,r} + \left[ A_{\bar{m}\bar{m}2}\delta(r-r(t)) \right]_{,rr}\Big\}\:,
\end{split}
\end{equation}
where $[\,\cdot\,]_{,r}$ denotes a derivative with respect to $r$, and $[\,\cdot\,]_{,rr}$ denotes a second derivative with respect to $r$, and
\begin{align}
    A_{nn0} &= -\frac{2}{\sqrt{2\pi}\Delta^2}C_{nn}\rho^{-2}\bar{\rho}^{-1}L^\dagger_1 \left[ \rho^{-4}L^\dagger_2 (\rho^3 S) \right]\:,\\
    A_{\bar{m}n0} &= \frac{2}{\sqrt{\pi}\Delta} C_{\bar{m}n} \rho^{-3} \left[ (L^\dagger_2 S)\left( \mathrm{i}\frac{K}{\Delta}+\rho+\bar{\rho} \right)- a \sin\theta S\frac{K}{\Delta}(\bar{\rho}-\rho) \right]\:,\\
    A_{\bar{m}\bar{m}0} &= -\frac{2}{\sqrt{2\pi}} \rho^{-3} \bar{\rho}~C_{\bar{m}\bar{m}} S  \left[-\mathrm{i}  \left(\frac{K}{\Delta}\right)_{,r} - \frac{K^2}{\Delta^2} + 2\mathrm{i} \rho \frac{K}{\Delta}\right]\:,\\
    A_{\bar{m}n1} &= \frac{2}{\sqrt{\pi}\Delta} \rho^{-3}C_{\bar{m}n} \left[ L^\dagger_2 S + \mathrm{i} a \sin\theta (\bar{\rho}-\rho)S\right]\:,\\
    A_{\bar{m}\bar{m}1} &= -\frac{2}{\sqrt{2\pi}} \rho^{-3} \bar{\rho}~C_{\bar{m}\bar{m}} S \left( \mathrm{i} \frac{K}{\Delta} + \rho \right)\:,\\
    A_{\bar{m}\bar{m}2} &= -\frac{1}{\sqrt{2\pi}} \rho^{-3} \bar{\rho}~C_{\bar{m}\bar{m}} S\:.
\end{align}

Substituting these expressions into Eq.~\eqref{eq:R_infty} and performing a partial integral, $\tilde{Z}^\mathrm{H}_{\ell m\omega}$ can be written as
\begin{equation}
    \tilde{Z}^\mathrm{H}_{\ell m\omega} = \frac{\mu}{2\mathrm{i}\omega B^\mathrm{inc}_{\ell m\omega}}\int^\infty_{\infty} \diff t~\mathrm{e}^{\mathrm{i}\omega t-\mathrm{i}m \phi(t)}W_{\ell m\omega}(r(t))\:,
\end{equation}
where 
\begin{equation}
    W_{\ell m\omega} = \left[ R^\mathrm{in}_{\ell m\omega} \left( A_{nn0} + A_{\bar{m}n0} + A_{\bar{m}\bar{m}0}\right) -\frac{\diff R^\mathrm{in}_{\ell m\omega}}{\diff r}\left( A_{\bar{m}n1} + A_{\bar{m}\bar{m}1} \right) + \frac{\diff^2 R^\mathrm{in}_{\ell m\omega}}{\diff r^2} A_{\bar{m}\bar{m}2} \right]_{r=r(t)}\:.
\end{equation}
\end{widetext}

\section{Sasaki-Nakamura equation of the Kerr--Newman BH}
\label{app:SN_eq}
In this appendix, we provide details of the Sasaki-Nakamura equation for the Kerr-Newman BH.
The Sasaki-Nakamura equation is constructed to have a short-ranged potential, specifically, reducing to the Regge-Wheeler equation in the limit of vanishing BH spin~\cite{Sasaki:1981sx}.
The new variable $X_{\ell m\omega}$ is related to $R_{\ell m\omega}$ by 
\begin{equation}
\label{eq:R_X}
\begin{split}
    R_{\ell m\omega} = \frac{1}{\gamma} \Bigg[ &\frac{\alpha \Delta + \beta'}{\sqrt{r^2+a^2}} X_{\ell m\omega}- \frac{\beta}{\Delta}\left(\frac{\Delta X_{\ell m\omega}}{\sqrt{r^2+a^2}}\right)' \Bigg]\:,    
\end{split}
\end{equation}
where
\begin{align}
    \alpha(\omega, r) &= 3\mathrm{i}K' + \lambda_{\ell m\omega} + \frac{6\Delta}{r^2} - \mathrm{i}\frac{K \beta}{\Delta^2}\:,\\
    \beta(\omega, r) &= \Delta \left(-2\mathrm{i}K + \Delta' - \frac{4\Delta}{r}\right)\:,\\
    \gamma(\omega, r) &= c_0(\omega) + \frac{c_1(\omega)}{r} + \frac{c_2(\omega)}{r^2} + \frac{c_3(\omega)}{r^3} + \frac{c_4(\omega)}{r^4} \label{eq:gamma}\:.
\end{align}
The coefficients $\{c_i(\omega)\}$ in Eq.~\eqref{eq:gamma} are given by
\begin{widetext}
\begin{align}
    c_0(\omega) &= \lambda(\lambda + 2) - 12 \mathrm{i} \omega M - 12 a \omega (a \omega - m), \\
    c_1(\omega) &= -8\mathrm{i} \left[ -a m \lambda - 3 Q^2 \omega + a^2 (-3 + \lambda) \omega \right], \\
    c_2(\omega) &= 12\left[ Q^2 + a \left( a - 2 a m^2 + 2 \mathrm{i} m M + 4 a^2 m \omega - 2 \mathrm{i} a M \omega - 2 a^3 \omega^2 \right) \right], \\
    c_3(\omega) &= 24\mathrm{i}(a^2 + Q^2) \left( -a m + \mathrm{i} M + a^2 \omega \right), \\
    c_4(\omega) &= 12 (a^2 + Q^2)^2.
\end{align}
\end{widetext}
The homogeneous Sasaki-Nakamura equation is expressed as
\begin{equation}
\label{eq:SN_eq}
    \left[ \frac{\diff^2}{\diff {r_\ast}^2} - F_{\ell m}(\omega, r)\frac{\diff}{\diff r_\ast} - U_{\ell m}(\omega,r) \right] X_{\ell m\omega} (r_\ast) = 0 \:,
\end{equation}
where
\begin{align}
    F_{\ell m}(\omega, r) &= \frac{\gamma'}{\gamma} \frac{\Delta}{r^2+a^2}\:,\\
    U_{\ell m}(\omega, r) &= \frac{\Delta U_1}{(r^2+a^2)^2} + G^2 + \frac{\Delta G'}{r^2+a^2} - FG\:, 
\end{align}
and
\begin{align}
    G(r) &= -\frac{2(r-M)}{r^2+a^2} + \frac{r\Delta}{(r^2+a^2)^2}\:, \\
    U_1(\omega, r) &= V + \frac{\Delta^2}{\beta}\left[ \left( 2\alpha + \frac{\beta'}{\Delta} \right)' - \frac{\gamma'}{\gamma}\left( \alpha + \frac{\beta'}{\Delta} \right) \right]\:.
\end{align}
Notice that the modification due to $Q$ does not appear in $c_{0}(\omega)$. 

Similar to Eqs.~\eqref{eq:in_T} and \eqref{eq:up_T}, we consider two linearly independent solutions to Eq.~\eqref{eq:SN_eq}, $X^\mathrm{in}_{\ell m\omega}$ and $X^\mathrm{up}_{\ell m\omega}$, whose asymptotic forms are given by
\begin{equation}
\label{eq:in_SN}
X^\mathrm{in}_{\ell m\omega}( r_\ast) \rightarrow \left\{
\begin{array}{ll}
A^\mathrm{trans}_{\ell m\omega}\mathrm{e}^{-\mathrm{i}kr_\ast}\:, & r_\ast \rightarrow  -\infty\\
A^\mathrm{ref}_{\ell m\omega}\mathrm{e}^{\mathrm{i}\omega r_\ast} + A^\mathrm{inc}_{\ell m\omega}\mathrm{e}^{-\mathrm{i}\omega r_\ast}\:, & r_\ast\rightarrow +\infty \\
\end{array}\;,
\right.
\end{equation}
\begin{equation}
\label{eq:up_SN}
X^\mathrm{up}_{\ell m\omega}( r_\ast) \rightarrow \left\{
\begin{array}{ll}
C^\mathrm{ref}_{\ell m\omega}\mathrm{e}^{-\mathrm{i}kr_\ast} + C^\mathrm{inc}_{\ell m\omega}\mathrm{e}^{\mathrm{i}kr_\ast}\:, & r_\ast \rightarrow  -\infty\\
C^\mathrm{trans}_{\ell m\omega}\mathrm{e}^{\mathrm{i}\omega r_\ast}\:, & r_\ast\rightarrow +\infty \\
\end{array}\:.
\right.
\end{equation}
Substituting these asymptotic behaviors into Eq.~\eqref{eq:R_X}, we obtain the relations between the coefficients $A$ and $B$ for the Teukolsky variables, and $C$ and $D$ for the Sasaki-Nakamura variables as
\begin{align}
    B^\mathrm{ref}_{\ell m\omega} &= -\frac{1}{4\omega^2} A^{\mathrm{ref}}_{\ell m\omega}\:, \\
    C^\mathrm{trans}_{\ell m\omega} &= \frac{4\omega^2}{c_0} D^\mathrm{trans}_{\ell m\omega}\:,\\
    B^\mathrm{trans}_{\ell m\omega} &= \frac{1}{d_{\ell m\omega}} A^\mathrm{trans}_{\ell m \omega}\:,
\end{align}
where
\begin{widetext}
\begin{equation}
d_{\ell m\omega}=4\sqrt{ r_{+}^{2}+a^{2}}~
\left[ 2\sqrt{M^2-(a^2+Q^2)} -\mathrm{i}k(r^2_+ + a^2) \right]\left[ \sqrt{M^2-(a^2+Q^2)}
- \mathrm{i}k(r^2_+ + a^2) \right]\:.
\end{equation}
\end{widetext}

\section{The Teukolsky-Starobinsky identity of the Kerr--Newman BH}
\label{app:TS_identity}
In this appendix, we generalize Eq.~(4.42) in Ref.~\cite{Teukolsky:1974yv} by keeping the function $\Delta(r)$ an arbitrary quadratic polynomial in $r$. 
In this case, the Teukolsky-Starobinsky identity between the Teukolsky radial functions for the spin weight $s=\pm 2$ denoted by $R_{\pm2}$ preserves its original form given by~\cite{Teukolsky:1974yv}
\begin{equation}
\label{eq:TS-id}
   4 \mathcal D \,  \mathcal D \, \mathcal D \, \mathcal D \, R_{-2} (r)= R_2 (r)\,,
\end{equation}
where
\begin{equation}
    \mathcal D := \frac{\diff}{\diff r} - \mathrm{i} \frac{K}{\Delta}\,.
\end{equation}
Plugging in\footnote{$C$ (denoted as $C_{\ell m \omega}$ in the main text) is a constant that is determined from the angular Teukolsky equation, which means $C$ should be unaffected if the angular Teukolsky equation is the same as Kerr in Ref.~\cite{Teukolsky:1974yv}.}
\begin{align}
    R_2 =  Y(r) \Delta^{-2} e^{-\mathrm{i} k r_*}\,, \quad 
    R_{-2} =  \frac{Z(r)}{C} \Delta^{2} e^{-\mathrm{i} k r_*}\,,
\end{align}
to Eq.~\eqref{eq:TS-id} and evaluating the equation at the outer event horizon $r=r_+$, we find
\begin{widetext}
\begin{align}
\label{eq:TS-id-gen}
    C Y(r_+) = 16 k \left(r_+^2+a^2\right) \left[4 \left(r_+^2+a^2\right)^2k^2+ \Delta '(r_+)^2\right] \left[
   \left(r_+^2+a^2\right)k+\mathrm{i} \Delta '(r_+)\right] Z(r_+) \,,
\end{align}
\end{widetext}
where a prime represents a radial derivative, and we used
\begin{equation}
    K(r_+) = (r_+^2+a^2)k\,. 
\end{equation}
Introducing further 
\begin{equation}
\label{eq:epsilon}
    \epsilon = \frac{\Delta '(r_+)}{4(r_+^2+a^2)}\,,
\end{equation}
Eq.~\eqref{eq:TS-id-gen} becomes
\begin{equation}
\label{eq:TS-id-gen-final}
     C Y(r_+) =64 k \left(r_+^2+a^2\right)^4 (k^2+4 \epsilon^2) (k+4 i \epsilon
   ) Z(r_+)\,.
\end{equation}
This is the generalized relation between $Y(r_+)$ and $Z(r_+)$ for an arbitrary quadratic function for $\Delta$.

Let us now comment on cases with specific forms of $\Delta$.
For the Kerr case, $r_+^2+a^2 = 2M r_+$, so the generalized relation in Eq.~\eqref{eq:TS-id-gen-final} reduces to
\begin{equation}
\label{eq:TS-id-Kerr}
     C Y(r_+) =64 k \left(2Mr_+\right)^4 (k^2+4 \epsilon_\mathrm{Kerr}^2) (k+4 \mathrm{i} \epsilon_\mathrm{Kerr}
   ) Z(r_+)\,,
\end{equation}
where 
\begin{equation}
    \epsilon_\mathrm{Kerr} = \frac{\sqrt{M^2-a^2}}{4Mr_+}\,,
\end{equation}
after using $\Delta '(r_+) = r_+ - r_- = 2 \sqrt{M^2-a^2}$,
in agreement with~\cite{Teukolsky:1974yv}.
For the Kerr--Newman case with $\Delta = r^2 - 2Mr + a^2 + Q^2$, 
\begin{equation}
    r_{\pm} = M \pm \sqrt{M^2 - (a^2+ Q^2)}\,,
\end{equation}
and the identity in Eq.~\eqref{eq:TS-id-gen-final} reduces to
\begin{equation}
\label{eq:TS-id-KN}
     C Y(r_+) =64 k \left(2Mr_+ - Q^2\right)^4 (k^2+4 \epsilon_\mathrm{KN}^2) (k+4 \mathrm{i} \epsilon_\mathrm{KN}
   ) Z(r_+)\,,
\end{equation}
with
\begin{equation}
    \epsilon_\mathrm{KN} = \frac{\sqrt{M^2 -(a^2+Q^2)}}{2(r_+^2+a^2)} = \frac{\sqrt{M^2 -(a^2+Q^2)}}{2(2Mr_+ - Q^2)}\,.
\end{equation}
Equation~\eqref{eq:TS-id-KN} for Kerr-Newman correctly reduces to Eq.~\eqref{eq:TS-id-Kerr} for Kerr in the limit $Q \to 0$.

\section{Validity check of the code}
\label{app:code_check}
\begin{figure}
    \centering
    \includegraphics[width=\linewidth]{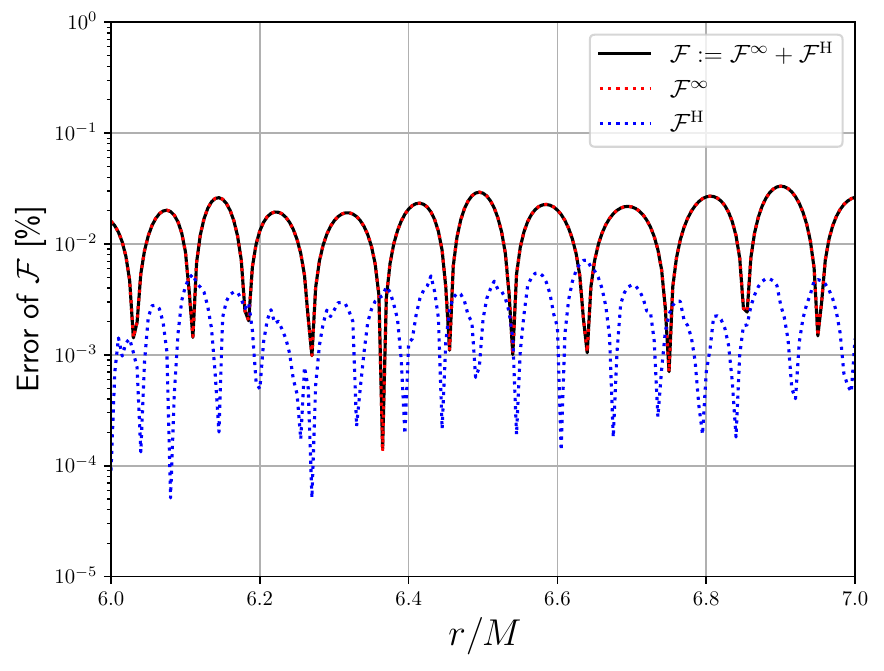}
    \caption{Relative error of $\mathcal{F}$ (black), $\mathcal{F}^\infty$ (red dotted), and $\mathcal{F}^\mathrm{H}$ (blue dotted) in the case of the Schwarzschild BH ($a/M=Q/M=0$) between the results using our code and \texttt{Black hole perturbation toolkit}~\cite{BHPToolkit}.}
    \label{fig:code_validity}
\end{figure}

For a validity check of our code, we compare $\mathcal{F}^{\infty,\mathrm{H}}$ derived using our code and \texttt{Black hole perturbation toolkit} (\texttt{BHPT})~\cite{BHPToolkit}.
We found that our result is consistent with the result of \texttt{BHPT} within the order of $10^{-2}~\%$ ($10^{-1}~\%$) for non-rotating (rotating) cases.
Figure~\ref{fig:code_validity} shows the error in the Schwarzschild case for $r_\mathrm{ISCO}\leq r \leq r_\mathrm{ISCO}+M$.

\section{Contour plots of the inspiral and post-inspiral SNRs}
\label{app:ins_pins_snrs}

\begin{figure*}
    \centering
    \includegraphics[width=0.95\linewidth]{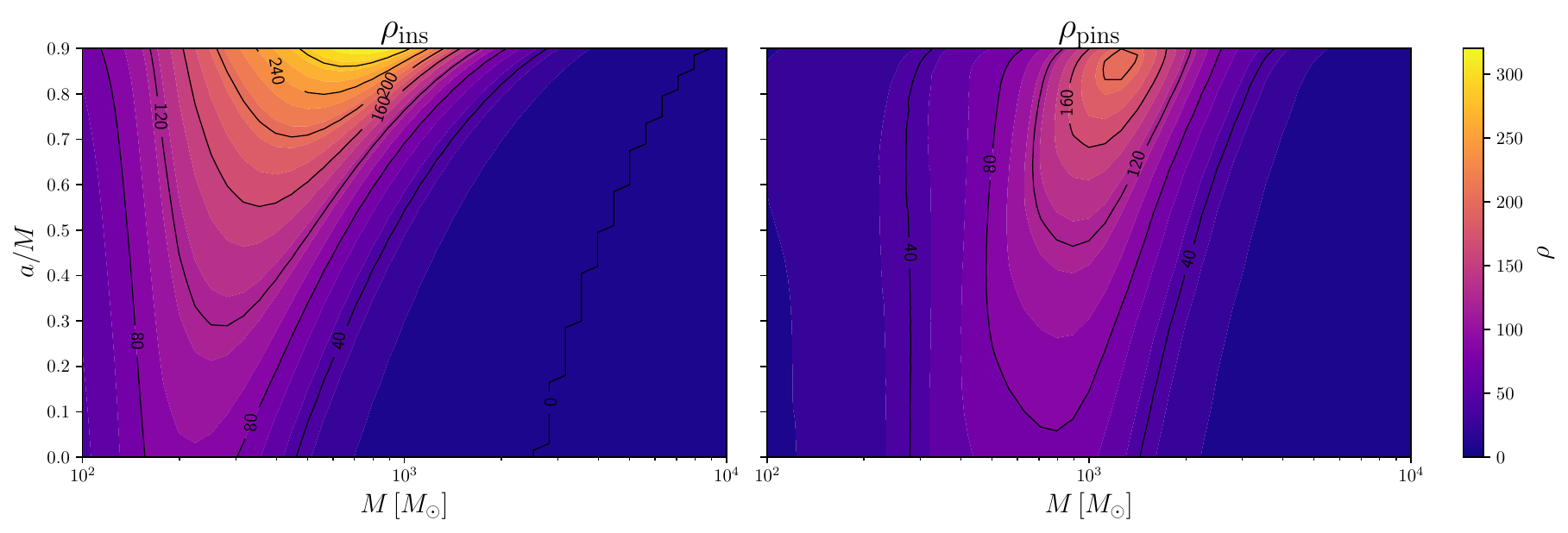}
    \caption{The sky-averaged inspiral (left) and post-inspiral (right) SNRs $\rho_\mathrm{ins}$ and $\rho_\mathrm{pins}$ for intermediate mass-ratio binary BH mergers observed with ET. 
    The color indicates the SNR as a function of $M$ and $a/M$, assuming fixed mass ratio $\eta=10^{-2}$ and luminosity distance $D=5~\mathrm{Gpc}$.}
    \label{fig:ins_post_ins_SNR}
\end{figure*}

Figure~\ref{fig:ins_post_ins_SNR} presents contour maps of the inspiral and post-inspiral SNRs, $\rho_\mathrm{ins}$ (left) and $\rho_\mathrm{pins}$ (right), for intermediate–mass–ratio BBH mergers observed with ET.  
Both SNRs are shown as functions of the primary mass $M$ and the dimensionless spin $a/M$, assuming a fixed mass ratio $\eta = 10^{-2}$ and luminosity distance $D =5~\mathrm{Gpc}$.

\newpage
\bibliographystyle{apsrev4-2}
\bibliography{reference}

@article{Apte:2019txp,
    author = "Apte, Anuj and Hughes, Scott A.",
    title = "{Exciting black hole modes via misaligned coalescences: I. Inspiral, transition, and plunge trajectories using a generalized Ori-Thorne procedure}",
    eprint = "1901.05901",
    archivePrefix = "arXiv",
    primaryClass = "gr-qc",
    doi = "10.1103/PhysRevD.100.084031",
    journal = "Phys. Rev. D",
    volume = "100",
    number = "8",
    pages = "084031",
    year = "2019"
}

@article{Geroch:1973am,
    author = "Geroch, Robert P. and Held, A. and Penrose, R.",
    title = "{A space-time calculus based on pairs of null directions}",
    doi = "10.1063/1.1666410",
    journal = "J. Math. Phys.",
    volume = "14",
    pages = "874--881",
    year = "1973"
}

@misc{BHPToolkit,
  title = {{Black Hole Perturbation Toolkit}},
  howpublished = {(\href{http://bhptoolkit.org/}{bhptoolkit.org})},
}

@article{Isaacson:1968zza,
    author = "Isaacson, Richard A.",
    title = "{Gravitational Radiation in the Limit of High Frequency. II. Nonlinear Terms and the Ef fective Stress Tensor}",
    doi = "10.1103/PhysRev.166.1272",
    journal = "Phys. Rev.",
    volume = "166",
    pages = "1272--1279",
    year = "1968"
}

@article{Teukolsky:1974yv,
    author = "Teukolsky, S. A. and Press, W. H.",
    title = "{Perturbations of a rotating black hole. III - Interaction of the hole with gravitational and electromagnetic radiation}",
    doi = "10.1086/153180",
    journal = "Astrophys. J.",
    volume = "193",
    pages = "443--461",
    year = "1974"
}

@article{Sasaki:1981sx,
    author = "Sasaki, Misao and Nakamura, Takashi",
    title = "{Gravitational Radiation From a Kerr Black Hole. 1. Formulation and a Method for Numerical Analysis}",
    reportNumber = "RIFP-461",
    doi = "10.1143/PTP.67.1788",
    journal = "Prog. Theor. Phys.",
    volume = "67",
    pages = "1788",
    year = "1982"
}

@article{Ori:2000zn,
    author = "Ori, Amos and Thorne, Kip S.",
    title = "{The Transition from inspiral to plunge for a compact body in a circular equatorial orbit around a massive, spinning black hole}",
    eprint = "gr-qc/0003032",
    archivePrefix = "arXiv",
    doi = "10.1103/PhysRevD.62.124022",
    journal = "Phys. Rev. D",
    volume = "62",
    pages = "124022",
    year = "2000"
}

@article{Stein:2019mop,
    author = "Stein, Leo C.",
    title = "{qnm: A Python package for calculating Kerr quasinormal modes, separation constants, and spherical-spheroidal mixing coefficients}",
    eprint = "1908.10377",
    archivePrefix = "arXiv",
    primaryClass = "gr-qc",
    doi = "10.21105/joss.01683",
    journal = "J. Open Source Softw.",
    volume = "4",
    number = "42",
    pages = "1683",
    year = "2019"
}

@article{Liu:2017fjx,
    author = "Liu, Chen-Yu and Lee, Da-Shin and Lin, Chi-Yong",
    title = "{Geodesic motion of neutral particles around a Kerr{\textendash}Newman black hole}",
    eprint = "1706.05466",
    archivePrefix = "arXiv",
    primaryClass = "gr-qc",
    doi = "10.1088/1361-6382/aa903b",
    journal = "Class. Quant. Grav.",
    volume = "34",
    number = "23",
    pages = "235008",
    year = "2017"
}

@misc{Khalvati:2025znb,
    author = "Khalvati, Hassan and Lynch, Philip and Burke, Ollie and Speri, Lorenzo and van de Meent, Maarten and Nasipak, Zachary",
    title = "{Systematic errors in fast relativistic waveforms for Extreme Mass Ratio Inspirals}",
    eprint = "2509.08875",
    archivePrefix = "arXiv",
    primaryClass = "gr-qc",
    month = "9",
    year = "2025"
}

@article{Bonga:2019ycj,
    author = "Bonga, B{\'e}atrice and Yang, Huan and Hughes, Scott A.",
    title = "{Tidal resonance in extreme mass-ratio inspirals}",
    eprint = "1905.00030",
    archivePrefix = "arXiv",
    primaryClass = "gr-qc",
    doi = "10.1103/PhysRevLett.123.101103",
    journal = "Phys. Rev. Lett.",
    volume = "123",
    number = "10",
    pages = "101103",
    year = "2019"
}

@misc{Babak:2021mhe,
    author = "Babak, Stanislav and Petiteau, Antoine and Hewitson, Martin",
    title = "{LISA Sensitivity and SNR Calculations}",
    eprint = "2108.01167",
    archivePrefix = "arXiv",
    primaryClass = "astro-ph.IM",
    reportNumber = "LISA-LCST-SGS-TN-001",
    month = "8",
    year = "2021"
}

@article{Planck:2015fie,
    author = "Ade, P. A. R. and others",
    collaboration = "Planck",
    title = "{Planck 2015 results. XIII. Cosmological parameters}",
    eprint = "1502.01589",
    archivePrefix = "arXiv",
    primaryClass = "astro-ph.CO",
    doi = "10.1051/0004-6361/201525830",
    journal = "Astron. Astrophys.",
    volume = "594",
    pages = "A13",
    year = "2016"
}

@article{Lim:2019xrb,
    author = "Lim, Halston and Khanna, Gaurav and Apte, Anuj and Hughes, Scott A.",
    title = "{Exciting black hole modes via misaligned coalescences: II. The mode content of late-time coalescence waveforms}",
    eprint = "1901.05902",
    archivePrefix = "arXiv",
    primaryClass = "gr-qc",
    doi = "10.1103/PhysRevD.100.084032",
    journal = "Phys. Rev. D",
    volume = "100",
    number = "8",
    pages = "084032",
    year = "2019"
}

@article{Bozzola:2020mjx,
    author = "Bozzola, Gabriele and Paschalidis, Vasileios",
    title = "{General Relativistic Simulations of the Quasicircular Inspiral and Merger of Charged Black Holes: GW150914 and Fundamental Physics Implications}",
    eprint = "2006.15764",
    archivePrefix = "arXiv",
    primaryClass = "gr-qc",
    doi = "10.1103/PhysRevLett.126.041103",
    journal = "Phys. Rev. Lett.",
    volume = "126",
    number = "4",
    pages = "041103",
    year = "2021"
}

@article{Wald:1974np,
    author = "Wald, Robert M.",
    title = "{Black hole in a uniform magnetic field}",
    doi = "10.1103/PhysRevD.10.1680",
    journal = "Phys. Rev. D",
    volume = "10",
    pages = "1680--1685",
    year = "1974"
}

@article{Moffat:2005si,
    author = "Moffat, J. W.",
    title = "{Scalar-tensor-vector gravity theory}",
    eprint = "gr-qc/0506021",
    archivePrefix = "arXiv",
    doi = "10.1088/1475-7516/2006/03/004",
    journal = "JCAP",
    volume = "03",
    pages = "004",
    year = "2006"
}

@article{Moffat:2016gkd,
    author = "Moffat, J. W.",
    title = "{LIGO GW150914 and GW151226 gravitational wave detection and generalized gravitation theory (MOG)}",
    eprint = "1603.05225",
    archivePrefix = "arXiv",
    primaryClass = "gr-qc",
    doi = "10.1016/j.physletb.2016.10.082",
    journal = "Phys. Lett. B",
    volume = "763",
    pages = "427--433",
    year = "2016"
}

@article{Cardoso:2016olt,
    author = "Cardoso, Vitor and Macedo, Caio F. B. and Pani, Paolo and Ferrari, Valeria",
    title = "{Black holes and gravitational waves in models of minicharged dark matter}",
    eprint = "1604.07845",
    archivePrefix = "arXiv",
    primaryClass = "hep-ph",
    doi = "10.1088/1475-7516/2016/05/054",
    journal = "JCAP",
    volume = "05",
    pages = "054",
    year = "2016",
    note = "[Erratum: JCAP 04, E01 (2020)]"
}

@article{LIGOScientific:2016aoc,
    author = "Abbott, B. P. and others",
    collaboration = "LIGO Scientific, Virgo",
    title = "{Observation of Gravitational Waves from a Binary Black Hole Merger}",
    eprint = "1602.03837",
    archivePrefix = "arXiv",
    primaryClass = "gr-qc",
    reportNumber = "LIGO-P150914",
    doi = "10.1103/PhysRevLett.116.061102",
    journal = "Phys. Rev. Lett.",
    volume = "116",
    number = "6",
    pages = "061102",
    year = "2016"
}

@article{LIGOScientific:2018mvr,
    author = "Abbott, B. P. and others",
    collaboration = "LIGO Scientific, Virgo",
    title = "{GWTC-1: A Gravitational-Wave Transient Catalog of Compact Binary Mergers Observed by LIGO and Virgo during the First and Second Observing Runs}",
    eprint = "1811.12907",
    archivePrefix = "arXiv",
    primaryClass = "astro-ph.HE",
    reportNumber = "LIGO-P1800307",
    doi = "10.1103/PhysRevX.9.031040",
    journal = "Phys. Rev. X",
    volume = "9",
    number = "3",
    pages = "031040",
    year = "2019"
}

@article{LIGOScientific:2020ibl,
    author = "Abbott, R. and others",
    collaboration = "LIGO Scientific, Virgo",
    title = "{GWTC-2: Compact Binary Coalescences Observed by LIGO and Virgo During the First Half of the Third Observing Run}",
    eprint = "2010.14527",
    archivePrefix = "arXiv",
    primaryClass = "gr-qc",
    reportNumber = "P2000061",
    doi = "10.1103/PhysRevX.11.021053",
    journal = "Phys. Rev. X",
    volume = "11",
    pages = "021053",
    year = "2021"
}

@article{KAGRA:2021vkt,
    author = "Abbott, R. and others",
    collaboration = "KAGRA, VIRGO, LIGO Scientific",
    title = "{GWTC-3: Compact Binary Coalescences Observed by LIGO and Virgo during the Second Part of the Third Observing Run}",
    eprint = "2111.03606",
    archivePrefix = "arXiv",
    primaryClass = "gr-qc",
    reportNumber = "LIGO-P2000318",
    doi = "10.1103/PhysRevX.13.041039",
    journal = "Phys. Rev. X",
    volume = "13",
    number = "4",
    pages = "041039",
    year = "2023"
}

@misc{LIGOScientific:2025slb,
    author = "Abac, A. G. and others",
    collaboration = "LIGO Scientific, VIRGO, KAGRA",
    title = "{GWTC-4.0: Updating the Gravitational-Wave Transient Catalog with Observations from the First Part of the Fourth LIGO-Virgo-KAGRA Observing Run}",
    eprint = "2508.18082",
    archivePrefix = "arXiv",
    primaryClass = "gr-qc",
    reportNumber = "LIGO-P2400386",
    month = "8",
    year = "2025"
}

@article{Carullo:2021oxn,
    author = "Carullo, Gregorio and Laghi, Danny and Johnson-McDaniel, Nathan K. and Del Pozzo, Walter and Dias, Oscar J. C. and Godazgar, Mahdi and Santos, Jorge E.",
    title = "{Constraints on Kerr-Newman black holes from merger-ringdown gravitational-wave observations}",
    eprint = "2109.13961",
    archivePrefix = "arXiv",
    primaryClass = "gr-qc",
    doi = "10.1103/PhysRevD.105.062009",
    journal = "Phys. Rev. D",
    volume = "105",
    number = "6",
    pages = "062009",
    year = "2022"
}

@article{Ackerman:2008kmp,
    author = "Ackerman, Lotty and Buckley, Matthew R. and Carroll, Sean M. and Kamionkowski, Marc",
    editor = "Klapdor-Kleingrothaus, Hans Volker and Krivosheina, Irina V.",
    title = "{Dark Matter and Dark Radiation}",
    eprint = "0810.5126",
    archivePrefix = "arXiv",
    primaryClass = "hep-ph",
    doi = "10.1103/PhysRevD.79.023519",
    journal = "Phys. Rev. D",
    volume = "79",
    pages = "023519",
    year = "2009"
}

@article{Feng:2009mn,
    author = "Feng, Jonathan L. and Kaplinghat, Manoj and Tu, Huitzu and Yu, Hai-Bo",
    title = "{Hidden Charged Dark Matter}",
    eprint = "0905.3039",
    archivePrefix = "arXiv",
    primaryClass = "hep-ph",
    reportNumber = "UCI-TR-2009-06",
    doi = "10.1088/1475-7516/2009/07/004",
    journal = "JCAP",
    volume = "07",
    pages = "004",
    year = "2009"
}

@article{Foot:2014uba,
    author = "Foot, R. and Vagnozzi, S.",
    title = "{Dissipative hidden sector dark matter}",
    eprint = "1409.7174",
    archivePrefix = "arXiv",
    primaryClass = "hep-ph",
    doi = "10.1103/PhysRevD.91.023512",
    journal = "Phys. Rev. D",
    volume = "91",
    pages = "023512",
    year = "2015"
}

@article{Foot:2014osa,
    author = "Foot, R. and Vagnozzi, S.",
    title = "{Diurnal modulation signal from dissipative hidden sector dark matter}",
    eprint = "1412.0762",
    archivePrefix = "arXiv",
    primaryClass = "hep-ph",
    doi = "10.1016/j.physletb.2015.06.063",
    journal = "Phys. Lett. B",
    volume = "748",
    pages = "61--66",
    year = "2015"
}

@article{Agrawal:2016quu,
    author = "Agrawal, Prateek and Cyr-Racine, Francis-Yan and Randall, Lisa and Scholtz, Jakub",
    title = "{Make Dark Matter Charged Again}",
    eprint = "1610.04611",
    archivePrefix = "arXiv",
    primaryClass = "hep-ph",
    doi = "10.1088/1475-7516/2017/05/022",
    journal = "JCAP",
    volume = "05",
    pages = "022",
    year = "2017"
}

@article{Punturo:2010zz,
    author = "Punturo, M. and others",
    editor = "Ricci, Fulvio",
    title = "{The Einstein Telescope: A third-generation gravitational wave observatory}",
    doi = "10.1088/0264-9381/27/19/194002",
    journal = "Class. Quant. Grav.",
    volume = "27",
    pages = "194002",
    year = "2010"
}

@article{Reitze:2019iox,
    author = "Reitze, David and others",
    title = "{Cosmic Explorer: The U.S. Contribution to Gravitational-Wave Astronomy beyond LIGO}",
    eprint = "1907.04833",
    archivePrefix = "arXiv",
    primaryClass = "astro-ph.IM",
    reportNumber = "LIGO-P1900316",
    journal = "Bull. Am. Astron. Soc.",
    volume = "51",
    number = "7",
    pages = "035",
    year = "2019"
}

@misc{Baker:2019nia,
    author = "Baker, John and others",
    title = "{The Laser Interferometer Space Antenna: Unveiling the Millihertz Gravitational Wave Sky}",
    eprint = "1907.06482",
    archivePrefix = "arXiv",
    primaryClass = "astro-ph.IM",
    reportNumber = "FERMILAB-PUB-19-436-A",
    month = "7",
    year = "2019"
}

@article{LIGOScientific:2016vlm,
    author = "Abbott, B. P. and others",
    collaboration = "LIGO Scientific, Virgo",
    title = "{Properties of the Binary Black Hole Merger GW150914}",
    eprint = "1602.03840",
    archivePrefix = "arXiv",
    primaryClass = "gr-qc",
    reportNumber = "LIGO-P1500218",
    doi = "10.1103/PhysRevLett.116.241102",
    journal = "Phys. Rev. Lett.",
    volume = "116",
    number = "24",
    pages = "241102",
    year = "2016"
}

@article{Ghosh:2016qgn,
    author = "Ghosh, Abhirup and others",
    title = "{Testing general relativity using golden black-hole binaries}",
    eprint = "1602.02453",
    archivePrefix = "arXiv",
    primaryClass = "gr-qc",
    reportNumber = "LIGO-P1500185-V10, ICTS-2016-1, LIGO-P1500185-V11",
    doi = "10.1103/PhysRevD.94.021101",
    journal = "Phys. Rev. D",
    volume = "94",
    number = "2",
    pages = "021101",
    year = "2016"
}

@article{Ghosh:2017gfp,
    author = "Ghosh, Abhirup and Johnson-Mcdaniel, Nathan K. and Ghosh, Archisman and Mishra, Chandra Kant and Ajith, Parameswaran and Del Pozzo, Walter and Berry, Christopher P. L. and Nielsen, Alex B. and London, Lionel",
    title = "{Testing general relativity using gravitational wave signals from the inspiral, merger and ringdown of binary black holes}",
    eprint = "1704.06784",
    archivePrefix = "arXiv",
    primaryClass = "gr-qc",
    reportNumber = "LIGO-P1700006, ICTS-2017-3",
    doi = "10.1088/1361-6382/aa972e",
    journal = "Class. Quant. Grav.",
    volume = "35",
    number = "1",
    pages = "014002",
    year = "2018"
}

@article{LIGOScientific:2019fpa,
    author = "Abbott, B. P. and others",
    collaboration = "LIGO Scientific, Virgo",
    title = "{Tests of General Relativity with the Binary Black Hole Signals from the LIGO-Virgo Catalog GWTC-1}",
    eprint = "1903.04467",
    archivePrefix = "arXiv",
    primaryClass = "gr-qc",
    reportNumber = "LIGO-P1800316",
    doi = "10.1103/PhysRevD.100.104036",
    journal = "Phys. Rev. D",
    volume = "100",
    number = "10",
    pages = "104036",
    year = "2019"
}

@article{LIGOScientific:2020tif,
    author = "Abbott, R. and others",
    collaboration = "LIGO Scientific, Virgo",
    title = "{Tests of general relativity with binary black holes from the second LIGO-Virgo gravitational-wave transient catalog}",
    eprint = "2010.14529",
    archivePrefix = "arXiv",
    primaryClass = "gr-qc",
    reportNumber = "LIGO-P2000091",
    doi = "10.1103/PhysRevD.103.122002",
    journal = "Phys. Rev. D",
    volume = "103",
    number = "12",
    pages = "122002",
    year = "2021"
}

@misc{LIGOScientific:2021sio,
  author = "Abbott, R. and others",
  collaboration = "LIGO Scientific, VIRGO, KAGRA",
  title         = {Tests of General Relativity with GWTC-3},
  eprint        = {2112.06861},
  archivePrefix = {arXiv},
  primaryClass  = {gr-qc},
  reportNumber  = {LIGO-P2100275},
  year          = {2021},
  month         = dec,
  note          = {arXiv:2112.06861}
}

@article{Madekar:2024zdj,
    author = "Madekar, Sakshi Satish and Johnson-McDaniel, Nathan K. and Gupta, Anuradha and Ghosh, Abhirup",
    title = "{A meta inspiral{\textendash}merger{\textendash}ringdown consistency test of general relativity with gravitational wave signals from compact binaries}",
    eprint = "2405.05884",
    archivePrefix = "arXiv",
    primaryClass = "gr-qc",
    doi = "10.1088/1361-6382/adf02a",
    journal = "Class. Quant. Grav.",
    volume = "42",
    number = "16",
    pages = "165011",
    year = "2025"
}

@article{Shaikh:2024wyn,
    author = "Shaikh, Md Arif and Bhat, Sajad A. and Kapadia, Shasvath J.",
    title = "{A study of the inspiral-merger-ringdown consistency test with gravitational-wave signals from compact binaries in eccentric orbits}",
    eprint = "2402.15110",
    archivePrefix = "arXiv",
    primaryClass = "gr-qc",
    doi = "10.1103/PhysRevD.110.024030",
    journal = "Phys. Rev. D",
    volume = "110",
    number = "2",
    pages = "024030",
    year = "2024"
}

@article{Newman:1965my,
    author = "Newman, E T. and Couch, E. and Chinnapared, K. and Exton, A. and Prakash, A. and Torrence, R.",
    title = "{Metric of a Rotating, Charged Mass}",
    doi = "10.1063/1.1704351",
    journal = "J. Math. Phys.",
    volume = "6",
    pages = "918--919",
    year = "1965"
}

@article{Dudley:1977zz,
    author = "Dudley, Alan L. and Finley, J. D.",
    title = "{Separation of Wave Equations for Perturbations of General Type-D Space-Times}",
    doi = "10.1103/PhysRevLett.38.1505",
    journal = "Phys. Rev. Lett.",
    volume = "38",
    pages = "1505--1508",
    year = "1977"
}

@article{Dudley:1978vd,
    author = "Dudley, Alan L. and Finley, III, J. D.",
    title = "{Covariant Perturbed Wave Equations in Arbitrary Type $D$ Backgrounds}",
    reportNumber = "Print-78-0590 (NEW MEXICO)",
    doi = "10.1063/1.524064",
    journal = "J. Math. Phys.",
    volume = "20",
    pages = "311",
    year = "1979"
}

@article{Dias:2021yju,
    author = "Dias, Oscar J. C. and Godazgar, Mahdi and Santos, Jorge E. and Carullo, Gregorio and Del Pozzo, Walter and Laghi, Danny",
    title = "{Eigenvalue repulsions in the quasinormal spectra of the Kerr-Newman black hole}",
    eprint = "2109.13949",
    archivePrefix = "arXiv",
    primaryClass = "gr-qc",
    doi = "10.1103/PhysRevD.105.084044",
    journal = "Phys. Rev. D",
    volume = "105",
    number = "8",
    pages = "084044",
    year = "2022"
}

@article{Teukolsky:1973ha,
    author = "Teukolsky, Saul A.",
    title = "{Perturbations of a rotating black hole. 1. Fundamental equations for gravitational electromagnetic and neutrino field perturbations}",
    doi = "10.1086/152444",
    journal = "Astrophys. J.",
    volume = "185",
    pages = "635--647",
    year = "1973"
}

@article{Press:1973zz,
    author = "Press, William H. and Teukolsky, Saul A.",
    title = "{Perturbations of a Rotating Black Hole. II. Dynamical Stability of the Kerr Metric}",
    doi = "10.1086/152445",
    journal = "Astrophys. J.",
    volume = "185",
    pages = "649--674",
    year = "1973"
}

@article{Rink:2024swg,
    author = "Rink, Katie and Bachhar, Ritesh and Islam, Tousif and Rifat, Nur E. M. and Gonzalez-Quesada, Kevin and Field, Scott E. and Khanna, Gaurav and Hughes, Scott A. and Varma, Vijay",
    title = "{Gravitational wave surrogate model for spinning, intermediate mass ratio binaries based on perturbation theory and numerical relativity}",
    eprint = "2407.18319",
    archivePrefix = "arXiv",
    primaryClass = "gr-qc",
    doi = "10.1103/PhysRevD.110.124069",
    journal = "Phys. Rev. D",
    volume = "110",
    number = "12",
    pages = "124069",
    year = "2024"
}

@article{Hild:2010id,
    author = "Hild, S. and others",
    title = "{Sensitivity Studies for Third-Generation Gravitational Wave Observatories}",
    eprint = "1012.0908",
    archivePrefix = "arXiv",
    primaryClass = "gr-qc",
    doi = "10.1088/0264-9381/28/9/094013",
    journal = "Class. Quant. Grav.",
    volume = "28",
    pages = "094013",
    year = "2011"
}

@article{Berti:2005eb,
    author = "Berti, Emanuele and Kokkotas, Kostas D.",
    title = "{Quasinormal modes of Kerr-Newman black holes: Coupling of electromagnetic and gravitational perturbations}",
    eprint = "gr-qc/0502065",
    archivePrefix = "arXiv",
    doi = "10.1103/PhysRevD.71.124008",
    journal = "Phys. Rev. D",
    volume = "71",
    pages = "124008",
    year = "2005"
}

@misc{Saha:2025nsg,
    author = "Saha, Sagnik and Silva, Hector O.",
    title = "{Quasinormal modes of Kerr-Newman black holes: revisiting the Dudley-Finley approximation}",
    eprint = "2510.05354",
    archivePrefix = "arXiv",
    primaryClass = "gr-qc",
    month = "10",
    year = "2025"
}

@article{Dias:2015wqa,
    author = "Dias, Oscar J. C. and Godazgar, Mahdi and Santos, Jorge E.",
    title = "{Linear Mode Stability of the Kerr-Newman Black Hole and Its Quasinormal Modes}",
    eprint = "1501.04625",
    archivePrefix = "arXiv",
    primaryClass = "gr-qc",
    doi = "10.1103/PhysRevLett.114.151101",
    journal = "Phys. Rev. Lett.",
    volume = "114",
    number = "15",
    pages = "151101",
    year = "2015"
}

@article{Dias:2022oqm,
    author = "Dias, Oscar J. C. and Godazgar, Mahdi and Santos, Jorge E.",
    title = "{Eigenvalue repulsions and quasinormal mode spectra of Kerr-Newman: an extended study}",
    eprint = "2205.13072",
    archivePrefix = "arXiv",
    primaryClass = "gr-qc",
    doi = "10.1007/JHEP07(2022)076",
    journal = "JHEP",
    volume = "07",
    pages = "076",
    year = "2022"
}

@article{Zi:2022hcc,
    author = "Zi, Tieguang and Zhou, Ziqi and Wang, Hai-Tian and Li, Peng-Cheng and Zhang, Jian-dong and Chen, Bin",
    title = "{Analytic kludge waveforms for extreme-mass-ratio inspirals of a charged object around a Kerr-Newman black hole}",
    eprint = "2205.00425",
    archivePrefix = "arXiv",
    primaryClass = "gr-qc",
    doi = "10.1103/PhysRevD.107.023005",
    journal = "Phys. Rev. D",
    volume = "107",
    number = "2",
    pages = "023005",
    year = "2023"
}

@article{Watarai:2024huy,
    author = "Watarai, Daiki",
    title = "{Ringdown of a postinnermost stable circular orbit of a rapidly spinning black hole: Mass ratio dependence of higher harmonic quasinormal mode excitation}",
    eprint = "2408.16747",
    archivePrefix = "arXiv",
    primaryClass = "gr-qc",
    reportNumber = "RESCEU-13/24",
    doi = "10.1103/PhysRevD.110.124029",
    journal = "Phys. Rev. D",
    volume = "110",
    number = "12",
    pages = "124029",
    year = "2024"
}

@misc{Watarai:2024vni,
    author = "Watarai, Daiki and Oshita, Naritaka and Tsuna, Daichi",
    title = "{Slowly decaying ringdown of a rapidly spinning black hole II: Inferring the masses and spins of supermassive black holes with LISA}",
    eprint = "2403.12380",
    archivePrefix = "arXiv",
    primaryClass = "gr-qc",
    reportNumber = "RESCEU-5/24, YITP-24-30, RIKEN-iTHEMS-Report-24",
    month = "3",
    year = "2024"
}

@misc{Yin:2025kls,
    author = "Yin, Yucheng and Lo, Rico K. L. and Chen, Xian",
    title = "{Gravitational radiation from Kerr black holes using the Sasaki-Nakamura formalism: waveforms and fluxes at infinity}",
    eprint = "2511.08673",
    archivePrefix = "arXiv",
    primaryClass = "gr-qc",
    month = "11",
    year = "2025"
}

@article{Lo:2023fvv,
    author = "Lo, Rico K. L.",
    title = "{Recipes for computing radiation from a Kerr black hole using a generalized Sasaki-Nakamura formalism: Homogeneous solutions}",
    eprint = "2306.16469",
    archivePrefix = "arXiv",
    primaryClass = "gr-qc",
    doi = "10.1103/PhysRevD.110.124070",
    journal = "Phys. Rev. D",
    volume = "110",
    number = "12",
    pages = "124070",
    year = "2024"
}

@misc{LIGOScientific:2025obp,
    collaboration = "LIGO Scientific, VIRGO, KAGRA",
    title = "{Black Hole Spectroscopy and Tests of General Relativity with GW250114}",
    eprint = "2509.08099",
    archivePrefix = "arXiv",
    primaryClass = "gr-qc",
    reportNumber = "LIGO P2500461",
    month = "9",
    year = "2025"
}

@article{KAGRA:2025oiz,
    author = "Abac, A. G. and others",
    collaboration = "KAGRA, Virgo, LIGO Scientific",
    title = "{GW250114: Testing Hawking{\textquoteright}s Area Law and the Kerr Nature of Black Holes}",
    eprint = "2509.08054",
    archivePrefix = "arXiv",
    primaryClass = "gr-qc",
    reportNumber = "LIGO-P2500421",
    doi = "10.1103/kw5g-d732",
    journal = "Phys. Rev. Lett.",
    volume = "135",
    number = "11",
    pages = "111403",
    year = "2025"
}

@article{Mandel:2014tca,
    author = "Mandel, Ilya and Berry, Christopher P. L. and Ohme, Frank and Fairhurst, Stephen and Farr, Will M.",
    title = "{Parameter estimation on compact binary coalescences with abruptly terminating gravitational waveforms}",
    eprint = "1404.2382",
    archivePrefix = "arXiv",
    primaryClass = "gr-qc",
    doi = "10.1088/0264-9381/31/15/155005",
    journal = "Class. Quant. Grav.",
    volume = "31",
    pages = "155005",
    year = "2014"
}

@misc{Roy:2025kra,
    author = {Roy, Ayush and K{\"u}chler, Lorenzo and Pound, Adam and Panosso Macedo, Rodrigo},
    title = "{Black hole mergers beyond general relativity: a self-force approach}",
    eprint = "2510.11793",
    archivePrefix = "arXiv",
    primaryClass = "gr-qc",
    month = "10",
    year = "2025"
}

@article{Yunes:2016jcc,
    author = "Yunes, Nicolas and Yagi, Kent and Pretorius, Frans",
    title = "{Theoretical Physics Implications of the Binary Black-Hole Mergers GW150914 and GW151226}",
    eprint = "1603.08955",
    archivePrefix = "arXiv",
    primaryClass = "gr-qc",
    doi = "10.1103/PhysRevD.94.084002",
    journal = "Phys. Rev. D",
    volume = "94",
    number = "8",
    pages = "084002",
    year = "2016"
}

@article{Yunes:2025xwp,
    author = "Yunes, Nicol{\'a}s and Siemens, Xavier and Yagi, Kent",
    title = "{Gravitational-wave tests of general relativity with ground-based detectors and pulsar-timing arrays}",
    doi = "10.1007/s41114-024-00054-9",
    journal = "Living Rev. Rel.",
    volume = "28",
    number = "1",
    pages = "3",
    year = "2025"
}

@misc{Gupta:2025utd,
    author = "Gupta, Anuradha",
    title = "{Ten years of extreme gravity tests of general theory of relativity with gravitational-wave observations}",
    eprint = "2511.15890",
    archivePrefix = "arXiv",
    primaryClass = "gr-qc",
    reportNumber = "LIGO-P2500706",
    month = "11",
    year = "2025"
}

@article{Chatziioannou:2017tdw,
    author = "Chatziioannou, Katerina and Klein, Antoine and Yunes, Nicol{\'a}s and Cornish, Neil",
    title = "{Constructing Gravitational Waves from Generic Spin-Precessing Compact Binary Inspirals}",
    eprint = "1703.03967",
    archivePrefix = "arXiv",
    primaryClass = "gr-qc",
    doi = "10.1103/PhysRevD.95.104004",
    journal = "Phys. Rev. D",
    volume = "95",
    number = "10",
    pages = "104004",
    year = "2017"
}

@article{Colleoni:2024knd,
    author = "Colleoni, Marta and Vidal, Felip A. Ramis and Garc{\'\i}a-Quir{\'o}s, Cecilio and Ak{\c{c}}ay, Sarp and Bera, Sayantani",
    title = "{Fast frequency-domain gravitational waveforms for precessing binaries with a new twist}",
    eprint = "2412.16721",
    archivePrefix = "arXiv",
    primaryClass = "gr-qc",
    doi = "10.1103/PhysRevD.111.104019",
    journal = "Phys. Rev. D",
    volume = "111",
    number = "10",
    pages = "104019",
    year = "2025"
}

@article{Pratten:2020ceb,
    author = "Pratten, Geraint and others",
    title = "{Computationally efficient models for the dominant and subdominant harmonic modes of precessing binary black holes}",
    eprint = "2004.06503",
    archivePrefix = "arXiv",
    primaryClass = "gr-qc",
    doi = "10.1103/PhysRevD.103.104056",
    journal = "Phys. Rev. D",
    volume = "103",
    number = "10",
    pages = "104056",
    year = "2021"
}

@article{Khan:2015jqa,
    author = {Khan, Sebastian and Husa, Sascha and Hannam, Mark and Ohme, Frank and P{\"u}rrer, Michael and Jim{\'e}nez Forteza, Xisco and Boh{\'e}, Alejandro},
    title = "{Frequency-domain gravitational waves from nonprecessing black-hole binaries. II. A phenomenological model for the advanced detector era}",
    eprint = "1508.07253",
    archivePrefix = "arXiv",
    primaryClass = "gr-qc",
    doi = "10.1103/PhysRevD.93.044007",
    journal = "Phys. Rev. D",
    volume = "93",
    number = "4",
    pages = "044007",
    year = "2016"
}

@article{Ajith:2007qp,
    author = "Ajith, Parameswaran and others",
    editor = "Krishnan, B. and Papa, M. A. and Schutz, Bernard F.",
    title = "{Phenomenological template family for black-hole coalescence waveforms}",
    eprint = "0704.3764",
    archivePrefix = "arXiv",
    primaryClass = "gr-qc",
    doi = "10.1088/0264-9381/24/19/S31",
    journal = "Class. Quant. Grav.",
    volume = "24",
    pages = "S689--S700",
    year = "2007"
}

@article{Buonanno:1998gg,
    author = "Buonanno, A. and Damour, T.",
    title = "{Effective one-body approach to general relativistic two-body dynamics}",
    eprint = "gr-qc/9811091",
    archivePrefix = "arXiv",
    reportNumber = "IHES-P-98-74",
    doi = "10.1103/PhysRevD.59.084006",
    journal = "Phys. Rev. D",
    volume = "59",
    pages = "084006",
    year = "1999"
}

@article{Bohe:2016gbl,
    author = "Boh{\'e}, Alejandro and others",
    title = "{Improved effective-one-body model of spinning, nonprecessing binary black holes for the era of gravitational-wave astrophysics with advanced detectors}",
    eprint = "1611.03703",
    archivePrefix = "arXiv",
    primaryClass = "gr-qc",
    reportNumber = "LIGO-P1600315",
    doi = "10.1103/PhysRevD.95.044028",
    journal = "Phys. Rev. D",
    volume = "95",
    number = "4",
    pages = "044028",
    year = "2017"
}

@article{Cotesta:2018fcv,
    author = "Cotesta, Roberto and Buonanno, Alessandra and Boh{\'e}, Alejandro and Taracchini, Andrea and Hinder, Ian and Ossokine, Serguei",
    title = "{Enriching the Symphony of Gravitational Waves from Binary Black Holes by Tuning Higher Harmonics}",
    eprint = "1803.10701",
    archivePrefix = "arXiv",
    primaryClass = "gr-qc",
    doi = "10.1103/PhysRevD.98.084028",
    journal = "Phys. Rev. D",
    volume = "98",
    number = "8",
    pages = "084028",
    year = "2018"
}

@article{Lindblom:2008cm,
    author = "Lindblom, Lee and Owen, Benjamin J. and Brown, Duncan A.",
    title = "{Model Waveform Accuracy Standards for Gravitational Wave Data Analysis}",
    eprint = "0809.3844",
    archivePrefix = "arXiv",
    primaryClass = "gr-qc",
    doi = "10.1103/PhysRevD.78.124020",
    journal = "Phys. Rev. D",
    volume = "78",
    pages = "124020",
    year = "2008"
}

@article{Burke:2023lno,
    author = "Burke, Ollie and Piovano, Gabriel Andres and Warburton, Niels and Lynch, Philip and Speri, Lorenzo and Kavanagh, Chris and Wardell, Barry and Pound, Adam and Durkan, Leanne and Miller, Jeremy",
    title = "{Assessing the importance of first postadiabatic terms for small-mass-ratio binaries}",
    eprint = "2310.08927",
    archivePrefix = "arXiv",
    primaryClass = "gr-qc",
    doi = "10.1103/PhysRevD.109.124048",
    journal = "Phys. Rev. D",
    volume = "109",
    number = "12",
    pages = "124048",
    year = "2024"
}

@article{Yunes:2009ke,
    author = "Yunes, Nicolas and Pretorius, Frans",
    title = "{Fundamental Theoretical Bias in Gravitational Wave Astrophysics and the Parameterized Post-Einsteinian Framework}",
    eprint = "0909.3328",
    archivePrefix = "arXiv",
    primaryClass = "gr-qc",
    doi = "10.1103/PhysRevD.80.122003",
    journal = "Phys. Rev. D",
    volume = "80",
    pages = "122003",
    year = "2009"
}

@article{Tahura:2018zuq,
    author = "Tahura, Sharaban and Yagi, Kent",
    title = "{Parameterized Post-Einsteinian Gravitational Waveforms in Various Modified Theories of Gravity}",
    eprint = "1809.00259",
    archivePrefix = "arXiv",
    primaryClass = "gr-qc",
    doi = "10.1103/PhysRevD.98.084042",
    journal = "Phys. Rev. D",
    volume = "98",
    number = "8",
    pages = "084042",
    year = "2018",
    note = "[Erratum: Phys.Rev.D 101, 109902 (2020)]"
}

@article{Cano:2024jkd,
    author = {Cano, Pablo A. and Capuano, Lodovico and Franchini, Nicola and Maenaut, Simon and V{\"o}lkel, Sebastian H.},
    title = "{Parametrized quasinormal mode framework for modified Teukolsky equations}",
    eprint = "2407.15947",
    archivePrefix = "arXiv",
    primaryClass = "gr-qc",
    doi = "10.1103/PhysRevD.110.104007",
    journal = "Phys. Rev. D",
    volume = "110",
    number = "10",
    pages = "104007",
    year = "2024"
}

@article{Chung:2024ira,
    author = "Chung, Adrian Ka-Wai and Yunes, Nicolas",
    title = "{Ringing Out General Relativity: Quasinormal Mode Frequencies for Black Holes of Any Spin in Modified Gravity}",
    eprint = "2405.12280",
    archivePrefix = "arXiv",
    primaryClass = "gr-qc",
    doi = "10.1103/PhysRevLett.133.181401",
    journal = "Phys. Rev. Lett.",
    volume = "133",
    number = "18",
    pages = "181401",
    year = "2024"
}

@article{Hirano:2024fgp,
    author = "Hirano, Shin'ichi and Kimura, Masashi and Yamaguchi, Masahide and Zhang, Jiale",
    title = "{Parametrized black hole quasinormal ringdown formalism for higher overtones}",
    eprint = "2404.09672",
    archivePrefix = "arXiv",
    primaryClass = "gr-qc",
    reportNumber = "RUP-24-6",
    doi = "10.1103/PhysRevD.110.024015",
    journal = "Phys. Rev. D",
    volume = "110",
    number = "2",
    pages = "024015",
    year = "2024"
}

@article{Sasaki:2003xr,
    author = "Sasaki, Misao and Tagoshi, Hideyuki",
    title = "{Analytic black hole perturbation approach to gravitational radiation}",
    eprint = "gr-qc/0306120",
    archivePrefix = "arXiv",
    doi = "10.12942/lrr-2003-6",
    journal = "Living Rev. Rel.",
    volume = "6",
    pages = "6",
    year = "2003"
}

\end{document}